\documentclass[10pt,journal,final,twoside,twocolumn]{IEEEtran}

\usepackage{amsmath}
\usepackage{amssymb}
\usepackage{amsfonts}
\usepackage{epsfig}
\usepackage{times}
\usepackage{cite}
\usepackage{caption3}
\usepackage{graphicx}
\usepackage{paralist}

\begin{document}

\title{An Iteratively Decodable Tensor Product Code with Application to Data Storage}
\author{Hakim~Alhussien,~\IEEEmembership{Member,~IEEE,}
        and~Jaekyun~Moon,~\IEEEmembership{Fellow,~IEEE}
\thanks{Manuscript received January 15, 2009; revised August 1,
2009.}
\thanks{This work was supported in part by the NSF Theoretical Foundation Grant 0728676.}
\thanks{Hakim Alhussien is with Link-A-Media Devices, Santa Clara, CA  95051,
USA (e-mail:hakima@link-a-media.com).}
\thanks{Jaekyun Moon is a Professor of Electrical Engineering at KAIST, Yuseong-gu, Daejeon, 305-701, Republic of
Korea (e-mail:jaemoon@ee.kaist.ac.kr).}}

\markboth{IEEE JOURNAL ON SELECTED AREAS IN COMMUNICATIONS,~VOL.~28,~No.~2,~FEBRUARY~2010}%
{ALHUSSIEN \MakeLowercase{and} MOON: AN ITERATIVELY DECODABLE
TENSOR PRODUCT CODE WITH APPLICATION TO DATA STORAGE}

\maketitle


\begin{abstract}
The error pattern correcting code (EPCC) can be constructed to
provide a syndrome decoding table targeting the dominant error
events of an inter-symbol interference channel at the output of
the Viterbi detector. For the size of the syndrome table to be
manageable and the list of possible error events to be reasonable
in size, the codeword length of EPCC needs to be short enough.
However, the rate of such a short length code will be too low for
hard drive applications. To accommodate the required large
redundancy, it is possible to record only a highly compressed
function of the parity bits of EPCC's tensor product with a symbol
correcting code. In this paper, we show that the proposed tensor
error-pattern correcting code (T-EPCC) is linear time encodable
and also devise a low-complexity soft iterative decoding algorithm
for EPCC's tensor product with $q$-ary LDPC (T-EPCC-$q$LDPC).
Simulation results show that T-EPCC-$q$LDPC achieves almost
similar performance to single-level $q$LDPC with a $1/2$ KB sector
at $50\%$ reduction in decoding complexity. Moreover, $1$ KB
T-EPCC-$q$LDPC surpasses the performance of $1/2$ KB single-level
$q$LDPC at the same decoder complexity.
\end{abstract}




 \section{Introduction} \label{Intro}

\PARstart The advent of high recording density enabling
technologies, pioneered by galloping improvements in head and
media design and manufacturing processes, has pushed for similar
advances in read channel design and error correction coding,
driving research efforts into developing
channel-capacity-approaching coding schemes based on soft
iterative decoding that are also implementation
friendly~\cite{HaoDec07,HaoMarch07}. Soft decodable error
correction codes (ECC), mainly low density parity check (LDPC)
codes, would eventually replace conventional Reed-Solomon (RS)
outer ECC, which despite its large minimum distance, possesses a
dense parity check matrix that does not lend itself easily to
powerful belief propagation (BP) decoding. There exists vast
literature on the various design aspects of LDPC coded systems for
magnetic recording applications. This includes code
construction~\cite{Varnica03,Sanka03,Vasic04,Liva08}, efficient
encoding~\cite{Yang04,Zongwang06}, decoder
optimization~\cite{Han06,Kurkoski03,Han08}, and performance
evaluation~\cite{Cideciyan02,Hongxin01,Kumar07}. In this work, we
propose an LDPC coded system optimized for the magnetic recording
channel that spans contributions in most of these areas.

The error-pattern correcting code (EPCC) is proposed
in~\cite{MoonICC05,intermag06,intermag07} motivated by the
well-known observation that the error rate at the channel detector
output of an ISI channel is dominated by a few specific known
error cluster patterns. This is due to the fact that the channel
output energies associated with these error patterns are smaller
than those of other patterns. A multiparity cyclic EPCC was first
described in\cite{intermag06}, with an RS outer ECC, possessing
distinct syndrome sets for all such dominant error patterns. To
reduce the code rate penalty, which is a severe SNR degradation in
recording applications, a method to increase the code rate was
introduced in~\cite{intermag07} that also improved EPCC's
algebraic single and multiple error-pattern correction capability.
In this method, the generator polynomial of the short base EPCC is
multiplied by a primitive polynomial that is not already a factor
of the generator polynomial. Also, the primitive polynomial degree
is chosen so as to achieve a certain desired codeword length.
Moreover,~\cite{intermag07} describes a Viterbi detection
postprocessor that provides error-event-reliability information
aiding syndrome-mapping of EPCC to improve its correction
accuracy. However, improving the EPCC code rate by extending its
codeword length increases the probability of multiple dominant
error patterns within the codeword, and this requires increasing
the size of the syndrome table considerably to maintain the same
correction power, which eventually results in prohibitive decoding
complexity.
To maintain correction power with a manageable size
syndrome decoding table,~\cite{Jih_Moon} discusses a more
efficient method based on a list decoding strategy that delivers
satisfactory sector error rate (SER) gain with an outer RS ECC.
Later, this list decoding scheme was formulated as a soft-input
soft-output block in~\cite{Hakim07} and utilized to enhance the
performance of turbo equalization based on convolutional codes
(CC). Nevertheless, the serial concatenation scheme that proved
successful with RS hard decoding and CC-based turbo equalization
does not work as well in serial concatenation of long-EPCC and
LDPC. The reason is that when the LDPC decoder fails, especially
in the water-fall region, the sector contains a large number of
multiple error occurrences. When many such error events occur in a
given EPCC codeword, decoding by any reasonable size list decoder
is formidable. Thus, an inner EPCC cannot in any capacity reduce
the SER of a serially concatenated outer LDPC. On the other hand,
if the EPCC codeword length is decreased substantially, then the
number of errors per codeword is reasonable, as long as the
overall code rate is somehow kept high. Here, the concept of
tensor product construction comes into play.

Tensor product parity codes (TPPC) were first proposed in
\cite{wolf65} as the null-space of the parity check matrix
resulting from the tensor product of two other parity check
matrices corresponding to a variety of code families. As a result,
based on the choice of the concatenated codes, TPPC would be
classified as an error correction code if constructed from two
ECCs, an error detection code (EDC) if constructed from two EDCs,
and an error location code (ELC) if constructed from an EDC and an
ECC in a special order. As a matter of fact, ELCs were introduced
earlier in \cite{wolf63} and their algebraic properties studied in
detail, but later incorporated in the unified theme of TPPCs in
\cite{wolf65}. Furthermore, a generalized class of hard-decodable
ELCs was later suggested for application in the magnetic recording
channel in \cite{Fahrner04}. In addition, TPPCs can be generalized
by combining a number of codes on various extension fields with
shorter binary codes. For this more general case, a decoding
algorithm was developed in~\cite{Imai81}. An ECC-type TPPC was
applied to longitudinal magnetic recording in~\cite{Chai06_L}, and
to perpendicular magnetic recording in~\cite{Chai06_P}.
In~\cite{Chai06_L}, a hard decodable tensor product code based on
a single parity code and a BCH code is proposed as an inner code
for RS. This code is suitable for low density longitudinal
recording channels for which dominant errors have odd weights,
such as  $\{+\}$ and $\{+,-,+\}$. Also,~\cite{Chai06_L} proposes
that the hard decoder passes the corrected parity bits to a
Viterbi detector reflecting channel states and parity code states
in order to compute the decoder output. Later,~\cite{Chai06_P}
presented two methods for combining a tensor-product single parity
code with a distance-enhancing constrained code. This code
combination achieved more satisfactory performance with RS as an
outer code in high density perpendicular recording channels.

Our goal in this work is to utilize the concept of tensor product
concatenation to construct high rate soft-decodable EPCCs on the
symbol-level of the outer ECC. The EPCC target error list is
matched to the dominant error events normally observed in high
density perpendicular recording channels. Since dominant error
events in perpendicular recording are not only of odd
weight~\cite{Cideciyan02}, this requires that our EPCC be a
multiparity code. However, in this case, a Viterbi
detector matching the
channel and parity will have prohibitive complexity. In spite of
this, the performance of the optimal decoder of the baseline
parity-coded channel can be approached by the low complexity
detection postprocessing technique in~\cite{Jih_Moon}. We also
present in detail a low complexity highly parallel soft decoder
for T-EPCC and show  that it achieves a better
performance-complexity tradeoff compared to conventional iterative
decoding schemes.

\subsection{Notations and Definitions} \label{notation}
    \begin{itemize}
    \item For a linear code $C:(n, k, p)$, $n$ denotes the
    codeword length, $k$ denotes the user data length, and
    $p=n-k$ denotes the number of code parity bits.
    \item For a certain parity check matrix $H$ corresponding to a linear code $\{C: Hc^t=0,\ \forall c \in C\}$,
    a syndrome $s$ is the range of a perturbation of a codeword $H(c+e)^t=s$. A signature refers to the
    range under $H$ for any bit block, not necessarily a codeword formed of data and parity bits.
    \item The multilevel log-likelihood ratio (mlLLR) of a random variable $\beta \in GF(q)$ corresponding
    to the p.m.f. (probability mass function) $p_i(\beta)=\mathrm{Pr}(\beta = i),
    \sum_{i=0}^{q-1}p_i(\beta)=1$, can be defined as: $\gamma(\beta=i)=\log(\frac{p_i(\beta)}{p_0(\beta)}),\gamma(\beta=0) = 0
    $.
    \item $[\textbf{x}]_i^j$ denotes a local segment $[x_i, x_{i+1},...,x_j]$ of the
    sequence $x_k$.
    \item The period of a generator polynomial on $GF(2)$ corresponding to a linear code is equal to the
    order of that polynomial, as defined in~\cite{Lidl94}. Also, for a syndrome
    set $\{s^{(i)}\}_{i=0}^{L-1}$ that corresponds to all the $L$ possible starting
    positions of an error event, the period $P$ is defined as the
    smallest integer such that $s^{(\rho+P)}=s^{(\rho)}$~\cite{Jih_Moon}.

    \item Assume $\alpha_L=\log(\alpha)$ and
    $\beta_L=\log(\beta)$, then $(\alpha+\beta)_L= \log(e^{\alpha_L}+e^{\beta_L})$. Define
    $max^*\left(\begin{array}{*{1}c} \alpha_L \\ \beta_L
    \end{array}\right)=(\alpha+\beta)_L=\max(\alpha_L,\beta_L)+\log(1+e^{-|\alpha_L-\beta_L|})$.
    Also, $max^*\left(\left\{ \gamma_k\right\}_{k=a}^{b}\right)$
    and $\mathop {max^*}\limits_{k=a}^{b}\left(\gamma_k\right)$ are two
    different representations of the recursive implementation of
    $max^*$ acting on the elements of the set $\{\gamma_k\}_{k=a}^{b}$.
\end{itemize}

\subsection{Acronyms}
    \begin{itemize}
    \item TPPC: Tensor Product Parity Code.
    \item $q$LDPC: $q$-ary Low Density Parity Check code.
    \item RS: Reed Solomon code.
    \item BCJR: Bahl-Cocke-Jelinek-Raviv .
    \item T-EPCC: Tensor product Error Pattern Correction Code.
    \item T-EPCC-$q$LDPC and T-EPCC-RS: Tensor product of EPCC and
    $q$LDPC or RS, respectively.
    \item LLR: Log-Likelihood Ratio.
    \item mlLLR: multi-level Log-Likelihood Ratio.
    \item ML: Maximum Likelihood.
    \item MAP: Maximum \emph{A Posteriori}.
    \item QC: Quasi-Cyclic.
    \item SPA: Sum-Product Algorithm.
    \end{itemize}

\section{Review of EPCC and the Tensor Product Coding Paradigm} \label{secIII}
In this section we give a brief review on the concept of EPCC,
including the design of two example codes that will be utilized
later in the simulation study. Also, we review the tensor product
coding paradigm and present an encoding method that allows for
EPCC-based linear-time-encodable TPPCs.
\begin{figure*}[t]
\centering
\includegraphics*[width=5in]{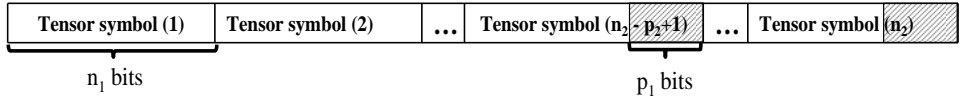}
\caption{TPPC $C_1(n_1,k_1,p_1) \otimes C_2(n_2,k_2,p_2)$ codeword
structure}\label{TPPCcodeword}
\end{figure*}
\subsection{EPCC Review and Examples}
We review constructing a cyclic code targeting the set of
$l_{max}$ dominant error events
$$\{e^{(1)}_k(x),e^{(2)}_k(x),...,e^{(l_{max})}_k(x)\}$$
represented as polynomials on $GF(2)$ that can occur at any
starting position $k$ in a codeword of length $l_T$. A syndrome of
error type $e^{(i)}(x)$ at position $k$ is defined as
$s^{(i)}_k(x) \equiv e^{(i)}_k(x)\ \mathbf{mod}\ g(x)$, where
$g(x)$ is the generator polynomial of the code and $\mathbf{mod}$
is the polynomial modulus operation.

A syndrome set $\mathbf{S}^{(i)}$ for error type $e^{(i)}(x)$
contains elements corresponding to all cyclic shifts of polynomial
$e^{(i)}(x)$; elements of $\mathbf{S}^{(i)}$ are thus related by
$s_{k+j}^{(i)} = x^{j}s_k^{(i)}\ \mathbf{mod}\ g(x)$.

For unambiguous decoding of $e^{(i)}(x)$ and $e^{(j)}(x)$,
$\forall\{i,j\}$, we must have $\mathbf{S}^{(i)} \cap
\mathbf{S}^{(j)} = \oslash$. This design requirement constrains
$g(x)$ to have distinct greatest common divisors with
$e^{(i)}(x)$, for all targeted $i$~\cite{intermag06}. However,
even if this constraint is satisfied, an element in
$\mathbf{S}^{(i)}$ can still map to more than one error position,
i.e., the period of the syndrome set- and period of $g(x)$- can be
less than $l_{T}$. Moreover, this constraint is only sufficient
but not necessary. As shown in~\cite{intermag06}, there may exist
a lower degree $g(x)$ that can yield distinct syndrome sets for
the targeted error polynomials, resulting in a higher rate EPCC. A
search method to find this $g(x)$ is already discussed in detail
in ~\cite{intermag06} and~\cite{Jih_Moon}.
 We next give two example EPCC constructions that will be used throughout the paper. We target the dominant error
events of the ``ideal" equalized monic channel $1+0.85D$ in AWGN,
which is suitable as a partial response target in perpendicular
magnetic recording read channels. For this channel, the dominant
errors are given by: $e^{(1)}(x)=1$, $e^{(2)}(x)=1+x$,
$e^{(3)}(x)=1+x+x^2$, etc., i.e. they can be represented as
polynomials on $GF(2)$ for which all powers of $x$ have nonzero
coefficients. The two EPCCs are:

\begin{itemize}
    \item \emph{Example 1}: Targeting error polynomials up to degree $4$, we get the generator polynomial $g(x)=1 +x
    +x^3 +x^5 +x^6$ of period $12$ via the search procedure of~\cite{intermag06}. Choosing a codeword length of
    $12$, $5$ distinct, non-overlapping syndrome sets are utilized to
    distinguish the $5$ target errors. Then, syndrome set $\mathbf{S}^{(3)}$ will have period $6$, while all other sets have period $12$. A syndrome set of period $6$ means that each syndrome decodes to one of $2$
    possible error positions within the $12$-bit codeword. Nonetheless, $e^{(3)}(x)$ can be decoded reliably via channel reliability
    information and the polarity of data support. The low code rate
    of $0.5$ makes this code unattractive as an inner code in a
    serial concatenation setup for recording channel applications.
    However, as we will see later, a tensor code setup makes it practical to use
such powerful codes for recording applications.
    \item \emph{Example 2}: Targeting error polynomials up to degree $9$, we have to record more redundancy. To accomplish this feat, a cyclic code with $8$ parity bits, code rate $0.56$, and a generator
    polynomial $g(x)=1 + x^2 +x^3 + x^5 + x^6 +x^8$ of period $18$
    is found by the search procedure in~\cite{intermag06}. Then, syndrome
    sets $\mathbf{S}^{(1)}$, $\mathbf{S}^{(3)}$, $\mathbf{S}^{(5)}$, and $\mathbf{S}^{(7)}$ each have period $18$ and thus can
    be decoded without ambiguity. While syndrome
    sets $\mathbf{S}^{(2)}$, $\mathbf{S}^{(4)}$, $\mathbf{S}^{(6)}$, $\mathbf{S}^{(8)}$, and $\mathbf{S}^{(10)}$ each have period $9$, decoding to one of two positions.
    The worst is $\mathbf{S}^{(9)}$ of period $2$, which would decode to one of $9$
    possible positions. Still, the algebraic decoder can quickly shrink this number to few positions by checking the data support,
    and then would choose the one position with highest local reliability.
\end{itemize}
\subsection{Tensor Product Parity Codes}\label{secIV}
\subsubsection{Construction and Properties of the TPPC Parity
Check Matrix} Consider a binary linear code $C_1:(n_1, k_1, p_1)$
derived from the null space of parity check matrix $H_{c_1}$, and
assume $C_1$ corrects any error event that belongs to class
$\varepsilon_1$. Also, consider a non-binary linear code
$C_2:(n_2, k_2, p_2)$ derived from the null space of parity check
matrix $H_{c_2}$ and defined over elements of $GF(2^{p_1})$.
Moreover, assume this code corrects any symbol error type that
belongs to class $\varepsilon_2$. As a preliminary step, convert
the binary $p_1 \times n_1$ matrix $H_{c_1}$, column by column,
into a string of $GF(2^{p_1})$ elements of dimension $1 \times
n_1$. Then, construct the matrix $$H_{c_3} = H_{c_1} \otimes
H_{c_2}$$ as a $p_2 \times n_1n_2$ array of $GF(2^{p_1})$
elements. Finally, convert the elements of $H_{c_3}$ into
$p_1$-bit columns, based on the same primitive polynomial of
degree $p_1$ used all over in the construction method. The null
space of the $p_1p_2 \times n_1n_2$ binary $H_{c_3}$ corresponds
to a linear binary code $C_3:(n_1n_2 ,k_3, p_1p_2)$. As shown in
Fig.~\ref{TPPCcodeword}, a $C_3$ codeword is composed of $n_2$
blocks termed ``tensor-symbols", each having $n_1$ bits. Also, it
can be shown that $C_3$ can correct any collection of tensor
symbol errors belonging to class $\varepsilon_2$, provided that
all errors within each tensor symbol belong to class
$\varepsilon_1$~\cite{wolf65}. Note that a tensor symbol is not an
actual $C_1$ codeword, and as such, using the terms ``inner" and
``outer" codes would not be completely accurate. In addition, the
tensor symbols are not codewords themselves, as can be seen in
Fig.~\ref{TPPCcodeword}, the first $k_2$ tensor symbols are all
data bits to start with, and even the last $p_2$ tensor symbols,
which are composed of data and parity bits, have non-zero
syndromes under $H_{c_1}$. Thus, a TPPC codeword does not
correspond directly to either $H_{c_1}$ or $H_{c_2}$, and as a
result, the component codebooks they describe are not recorded
directly on the channel. Another interesting property of the
resulting TPPC is that the symbol-mapping of the sequence of
tensor-symbol signatures under $H_{c_1}$ forms a codeword of
$C_2$, which we refer to as the ``signature-correcting component
code".
\begin{figure*}[t]
\centering
\includegraphics*[width=4.5in]{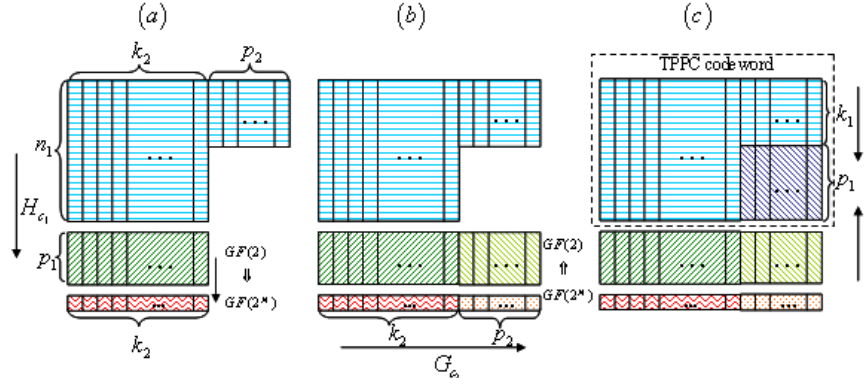}
\caption{TPPC encoder of $C_1(n_1,k_1,p_1) \otimes
C_2(n_2,k_2,p_2)$:(a) signatures calculated under $H_{c_1}$, then
the $p_1$-bit signatures are mapped to $GF(2^{p_1})$ symbols, (b)
$k_2$ signatures encoded by generator matrix $G_{c_2}$ into a
$C_2$ codeword, then parity symbols are mapped back to $GF(2)$,
(c) $p_1 \times p_2$ TPPC parity bits are calculated by back
substitution. }\label{TPPCencoder}
\end{figure*}
\subsubsection{Encoding of Tensor Product Parity Codes}

The encoding of a TPPC can be performed using its binary parity
check matrix, but the corresponding binary generator matrix is not
guaranteed to possess algebraic properties that enable linear time
encodability. Thus, an implementation-friendly approach would be
to utilize the encoders of the constituent codes, which can be
chosen to be linear time encodable.

Consider a binary code $C_1:(n_1, k_1, p_1)$ that is the null
space of parity check matrix $H_{c_1}$, and a non-binary code
$C_2:(n_2 ,k_2, p_2)$ defined on $GF(2^{p_1})$, the tensor-product
concatenation is a binary $C_3:(n_3 ,k_3, p_3)$, where: $$n_3 =
n_2 \times n_1, k_3 = n_1 \times n_2 - p_1 \times p_2, p_3 = p_1
\times p_2$$ Assume that $C_1$ is a cyclic code, and $C_2$ is any
of the linear time encodable codes, where we choose a quasi-cyclic
(QC) component code for the purpose of this study. Then, the
encoders of $C_1$ and $C_2$ communicate via the following
algorithm to generate a codeword of $C_3$, see
Fig.~\ref{TPPCencoder}:
\renewcommand{\labelenumi}{(\roman{enumi})}
\begin{enumerate}
    \item Receive a block of $n_1 \times k_2 + k_1 \times n_2 - k_1 \times k_2$ bits from the data source, call it major block $\alpha$.
    \item Divide major block $\alpha$ into minor block $\beta$ of $n_1 \times k_2$ bits, and minor block $\gamma$ of $k_1 \times n_2 - k_1 \times k_2$ bits
    (i.e. $k_1 \times p_2$ bits).
    \item Divide block $\beta$ into $k_2$ columns each of $n_1$ bits. Then, for each column, calculate the
    intermediate $p_1$-bit signature under the parity check matrix of
    $C_1$. Using a feedback shift register (FSR) to calculate the signatures, the computational cost is $\propto n_1$ operations per signature, and $\propto n_1 \times
    k_2$ for this entire step.
    \item Convert intermediate signatures from $p_1$-bit strings
    into $GF(2^{p_1})$ symbols.
    \item Encode the $k_2$ non-binary signatures into a $C_2$
    codeword of length $n_2$. Using FSRs to encode the quasi-cyclic $C_2$, the computational complexity of this step is $\propto n_2$.
    \item\label{step6} Convert computed signatures back into $p_1$-bit strings.
    \item Divide block $\gamma$ into $p_2$ columns each of $k_1$
    bits. Add $p_1$ blanks in each column to be filled with the parity
    bits of $C_3$. Then, align each column with the $p_2$ signatures computed in the
    previous step, leaving $p_1$ blanks in each column.
    \item Fill blanks in the previous step such that the signature
    of data plus parity blanks under $C_1$ equals the corresponding aligned
    signature from step (vi). The parity can be calculated using the systematic
    $H_{c_1}$ and the method of back substitution which requires a computational complexity
    $\propto n_1$ per column.
\end{enumerate}

The total computational complexity of this encoding algorithm is
$\propto n_1 \times k_2 + n_2 +n_1 \times p_2$, i.e. it is
$\propto n_1 \times n_2 = n_3$, which is the TPPC codeword length.
Thus, we have shown- with some constraints- that if $C_1$ and
$C_2$ are linear time encodable, then $C_3 = C_1 \otimes C_2$ is
linear time encodable.

\section{T-EPCC-RS Codes} \label{secV}
To demonstrate the algebraic properties of TPPC codes, we present
an example code suitable for recording applications with $1/2$ KB
sector size. Consider two component codes:
\begin{itemize}
    \item A binary cyclic $(18,10)$ EPCC of example 2 above with rate $0.556$, $8$
    parity bits, and parity check matrix in $GF(2^8)$:
\begin{equation}\label{eq5}
H_{epcc} = {\begin{array}{*{20}c}
   [1 & \alpha  & {\alpha ^2 } & {\alpha ^3 } & {\alpha ^4 } & {\alpha ^5 }
   & \ldots \\
    {\alpha ^6 } & {\alpha ^7 } & {\alpha ^{133} } & {\alpha ^{134} } & {\alpha ^{96} } & {\alpha ^{90} } & \ldots \\
    {\alpha ^{82} } & {\alpha ^{236} } & {\alpha ^{234} } & {\alpha ^{217} } & {\alpha ^{92} } & {\alpha ^{93} } & ]_{1 \times 18}.  \\
\end{array}}
\end{equation}
    \item A $(255,195)$ RS over $GF(2^8)$, of rate $0.765$, $t=30$, and $60$ parity
    symbols.
\end{itemize}
The resulting TPPC is a binary $(4590,4110)$ code, of rate
$0.896$, and redundancy of $480$ parity bits. For this code, a
codeword is made of $255$ $18$-bit tensor symbols, of which, any
combination of $30$ or less tensor symbol errors are correctable,
provided that each $18$-bit tensor symbol has a single or multiple
occurrence of a dominant error that is correctable by EPCC, those
being combinations of error polynomials up to degree $9$.
Furthermore, although the EPCC constituent code has a very low
rate of $0.556$, the resulting T-EPCC has a high rate of $0.895$.
Notably, in the view of the $18$-bit EPCC, this $61\%$ reduction
in recorded redundancy corresponds to an SNR improvement of $2$ dB
in a channel with rate penalty $\sim10\log_{10}(1/R)$, and $4.1$
dB in a channel with rate penalty $\sim10\log_{10}(1/R^2)$.
\begin{figure*}[t]
\centering
\includegraphics*[width=4.5in]{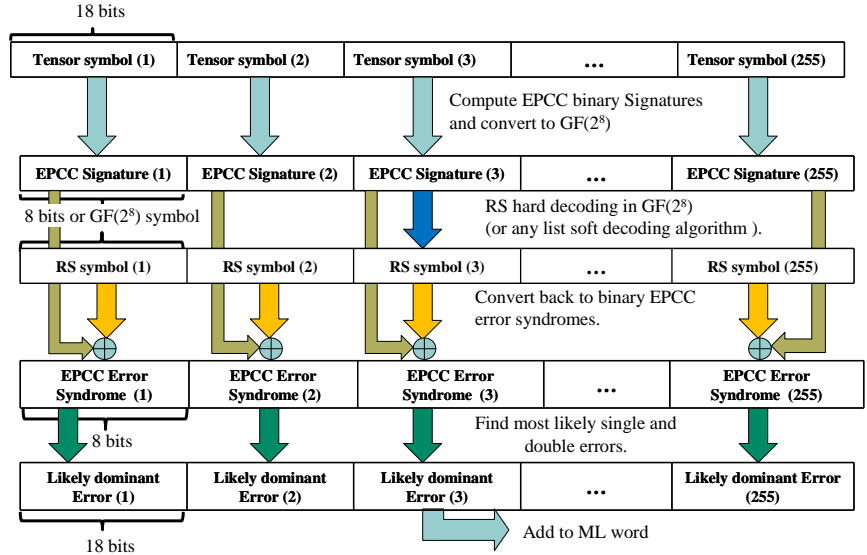}
\caption{Hard decoder of $(18,10,8)$ EPCC $\otimes$ $(255,195,2t)$
RS of $t=30$ over $GF(2^8)$.}\label{harddecoder}
\end{figure*}
\subsection{Hard Decoding of T-EPCC-RS Codes} \label{secVI}

Hard decoding of T-EPCC-RS directly reflects the code's algebraic
properties, and thus, serves to further clarify the concept of
tensor product codes. Hence, we discuss the hard decoding approach
before going into the design of soft decoding of T-EPCC codes. The
decoding algorithm is summarized by the following procedure, see
Fig.~\ref{harddecoder}:
\renewcommand{\labelenumi}{(\roman{enumi})}
\begin{enumerate}
    \item After hard slicing the output of the Viterbi channel detector, the signature of each tensor symbol is calculated under
    $H_{epcc}$. Each signature is then mapped into a Galois field symbol, where the sequence of non-binary signatures constitute
     an RS codeword - that is if the channel detector did not suffer any errors.
    \item Any hard-input RS decoder, such as the Berlekamp-Massey decoder, acts to find a legitimate RS codeword based on
     the observed signature-sequence.
    \item If the number of signature-symbols in error is larger than the RS correction power, RS decoding fails and the tensor product decoder halts.
    \item Otherwise, if RS decoding is deemed successful, the corrected signature-symbol sequence is added to the original observed signature-symbol sequence to generate the ``error syndrome-symbol" sequence.
    \item Each error syndrome-symbol is mapped into an EPCC bit-syndrome of the corresponding tensor
          symbol.
    \item Finally, EPCC decodes each tensor symbol to satisfy the error-syndrome generated by the component RS, in which it faces two
    scenarios:
    \begin{itemize}
        \item A zero ``error-syndrome" at the output of RS decoding indicates either no error occurred or a multiple error occurrence that has a zero EPCC-syndrome, which goes undetected.
        In this case, the EPCC decoder is turned off to save power.
        \item A non zero ``error-syndrome"  will turn EPCC correction
        on. If the error-syndrome indicates a single error occurrence in the target set, then, the EPCC
        single error algebraic decoder is turned on. On the other hand, if the error-syndrome is not recognized, then EPCC list decoding is turned on with a
        reasonable-size list of test words.
        \end{itemize}
        Note that although the number of  EPCC codewords (tensor symbols) is
        huge, the decoder complexity is reasonable since EPCC decoding is
        turned on only for nonzero error-syndromes.
\end{enumerate}

\section{T-EPCC-$q$LDPC Codes} \label{secVII}

We learned from the design of T-EPCC-RS that the component
signature-correcting codeword length can be substantially shorter
than the competing single level code. Although the minimum
distance is bound to be hurt if the increased redundancy does not
compensate for the shorter codeword length, employing iterative
soft decoding of the component signature-correcting code can
recover performance if designed properly. While LDPC codes have
strictly lower minimum distances compared to comparable rate and
code length RS codes, the sparsity of its parity check matrix
allows for effective belief propagation (BP) decoding. BP decoding
of LDPC codes consistently performs better than the best known
soft decoding algorithm for RS codes. Since the TPPC expansion
enables the use of $2$ to $4$ times shorter component LDPC
compared to a competing single level LDPC, a class of LDPC codes
efficient at such short lengths are critical. LDPC codes on high
order fields represent such good candidates. In that
respect,~\cite{Davey98} showed that the performance of binary LDPC
codes in AWGN can be significantly enhanced by a move to fields of
higher orders (extensions of $GF(2)$ being an example).
Moreover,~\cite{Davey98} established that for a monotonic
improvement in waterfall performance with field order, the parity
check matrix for very short blocks has to be very sparse.
Specifically, column weight $3$ codes over $GF(q)$ exhibit worse
bit-error-rate (BER) as $q$ increases, whereas column weight $2$
codes over $GF(q)$ exhibit monotonically lower BER as $q$
increases. These results were later confirmed in~\cite{Hu05},
where they also showed through a density evolution study of large
$q$ codes that optimum degree sequences favor a regular graph of
degree $2$ in all symbol nodes. On the other hand, for
satisfactory error floor performance, we found that using a column
weight higher than $2$ was necessary. This becomes more important
as the minimum distance decreases for lower $q$. For instance, we
found that a column weight of $3$ improved the error floor
behavior of $GF(2^6)$-LDPC at the expense of performance
degradation in the waterfall region.

\subsection{Design and Construction of $q$LDPC} The low rate and relatively low column weight design of $q$LDPC in a TPPC results in
a very sparse parity check matrix, allowing the usage of high
girth component $q$LDPC codes. To optimize the girth for a given
rate, we employ the progressive edge growth (PEG)
algorithm~\cite{Hu05} in $q$LDPC code design. PEG optimizes the
placement of a new edge connecting a particular symbol node to a
check node on the Tanner graph, such that the largest possible
local girth is achieved. Furthermore, PEG construction is very
flexible, allowing arbitrary code rates, Galois field sizes, and
column weights. In addition, modified PEG-construction with
linear-time encoding can be achieved without noticeable
performance degradation, facilitating the design of linear time
encodable tensor product codes. Of the two approaches to achieve
linear time encodability, namely, the upper triangular parity
check matrix construction~\cite{Hu05} and PEG construction with a
QC constraint~\cite{Li04}, we choose the latter approach, for
which the designed codes have better error floor behavior.
T-EPCC-$q$LDPC lends itself to iterative soft decoding quite
naturally. Next, we present a low complexity soft decoder
utilizing this important feature.

\begin{figure*}[t]
\centering
\includegraphics*[width=5in]{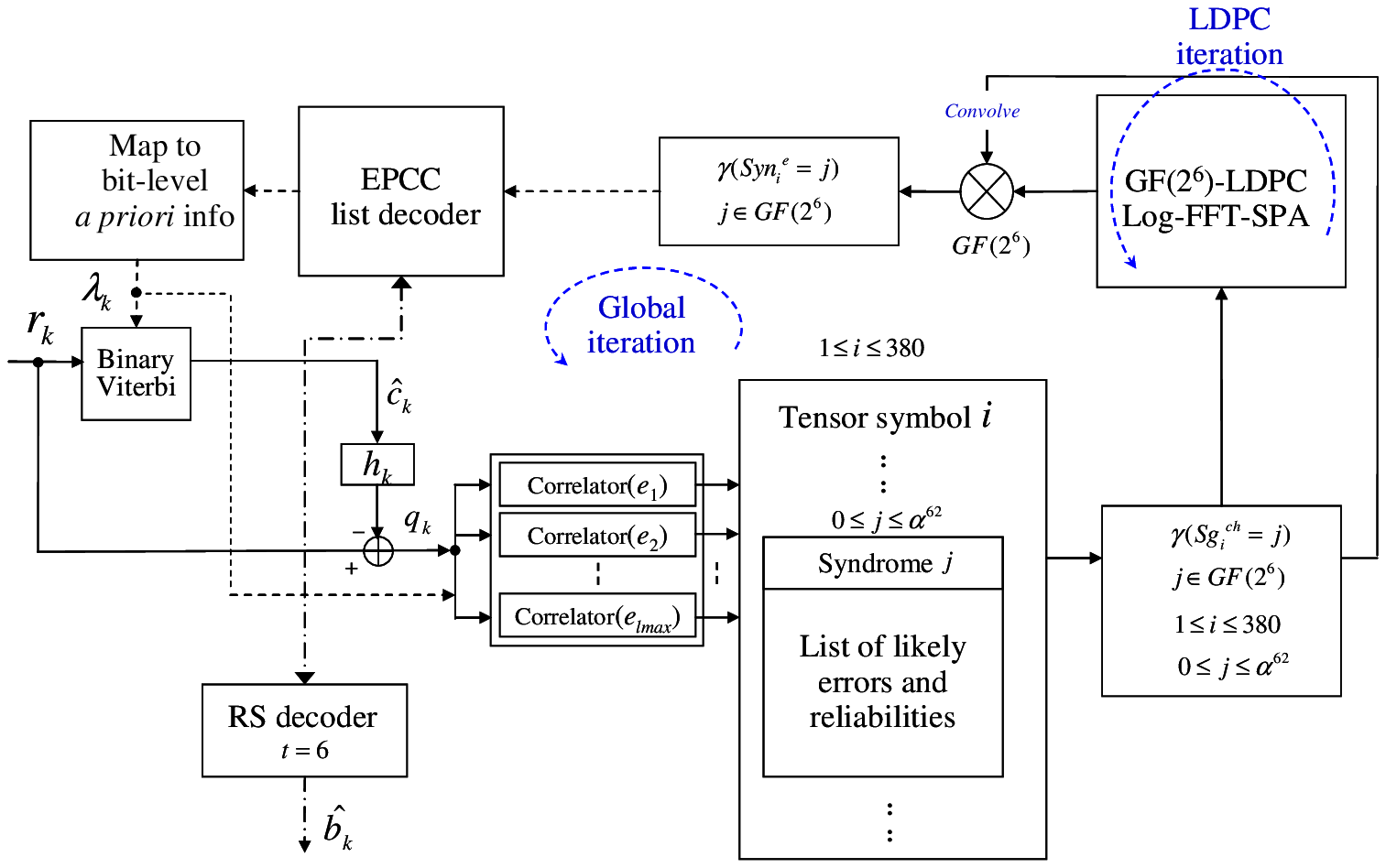}
\caption{T-EPCC-$q$LDPC soft decoder of $(12,6,6)$ EPCC $\otimes$
$(380,\ 323)$ $GF(2^6)$-LDPC.}\label{softDec}
\end{figure*}
\subsection{Soft Decoding of T-EPCC-$q$LDPC}\label{secVIII}
To fully utilize the power of the component codes in
T-EPCC-$q$LDPC, we need to develop a soft iterative version of the
hard decoder of T-EPCC-RS. To limit the complexity of the proposed
soft decoder, sub-optimal detection post-processing is adopted
instead of the maximum \emph{a posteriori} (MAP) detector to
evaluate tensor symbol signature reliabilities. The complexity of
the optimal MAP detector matched to both the channel of memory
length $L$ and $H_{epcc}$ of row length $p$ is exponential in
$p+L-1$. We present a practical soft detection scheme that
separates soft channel detection from tensor symbol signature
detection, though, through a component signature-correcting LDPC
in a TPPC setup, approaches the joint MAP performance through
channel iterations. The main stages of the decoder are, see
Fig.~\ref{softDec}:
\begin{enumerate}
\renewcommand{\labelenumi}{(\arabic{enumi})}
    \item \textbf{Detection postprocessing}:
        \begin{itemize}
            \item Utilizing \textit{a priori} information from the previous decoding iteration, binary Viterbi generates the hard ML word based on channel observations, for which the error sequence is calculated and passed to the correlator
          bank.
            \item A bank of local correlators estimates the probability of dominant error type/location pairs for all positions inside each tensor symbol.
        \end{itemize}
    \item \textbf{Signature p.m.f. calculation}:
    \begin{itemize}
        \item For each tensor symbol, the list of most likely error patterns is constructed. This list includes single occurrences and a predetermined set of their combinations.
              The list is then divided into sublists, each under the signature value it satisfies.
        \item For each tensor symbol, using each signature value's error likelihood list, we find the signature p.m.f. of that symbol.
    \end{itemize}

    \item \textbf{$q$-ary LDPC decoding}:
    \begin{itemize}
        \item Using the observed sequence of signature p.m.f.'s, we decode the component $q$-ary LDPC via FFT-based SPA.
        \item For each tensor symbol, the LDPC-corrected signature p.m.f. is convolved with the
        observed signature p.m.f. at its input to generate the error-syndrome p.m.f..
    \end{itemize}

    \item \textbf{EPCC decoding}:
    \begin{itemize}
        \item For each tensor symbol, we find the list of most probable
        error-syndromes and generate a list of test error words to satisfy each syndrome in the list.
        \item A bank of parallel EPCC single-error correcting decoders generates a list of most probable codewords along with their reliabilities.
    \end{itemize}

    \item \textbf{Bit-LLR feedback}:
    \begin{itemize}
        \item Using the codeword reliabilities we generate bit-level reliabilities that are fed back to the Viterbi detector and the detection postprocessing
        stage. Those bit-level reliabilities, serving as \textit{a priori} information, favor paths which satisfy both the ISI and parity constraints.
    \end{itemize}
\end{enumerate}
We explain each of these steps in the following sections, but we
replace any occurrence in the text of syndrome (signature) p.m.f.
by syndrome (signature) multi-level log-likelihood ratios (mlLLR),
as decoding will be entirely in the log domain for reasons
explained below.
\begin{figure}[h]
\centering
\includegraphics*[width=3.5in]{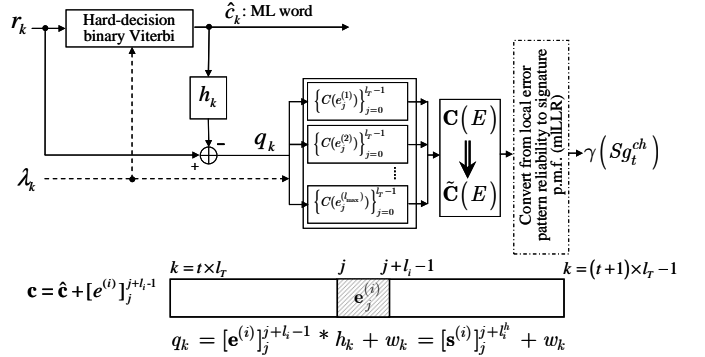}
\caption{Bank of parallel error-matched correlators to find error
pattern type/position reliabilities.} \label{Fig:Correlator}
\end{figure}
\subsubsection{Detection Postprocessing}
At this decoder stage we prepare a reliability matrix
${\bf{C}}\left( E \right)$ for error type/position pairs -
captured in a tensor symbol of length $l_T$- that is usable by the
next stage to calculate the tensor symbol's signature mlLLR:
\begin{eqnarray}\label{eq:C}
 \nonumber     \begin{array}{*{21}c}
                {\bf{C}}\left( E \right) = & & & & & & & & & & & &
                & & & & & & & &
               \end{array} & &  \\
   \nonumber \begin{array}{*{20}c}
   {} &
   \begin{array}{*{20}c}
    0  & & & & 1 & & & & \cdots & & & l_T \\
   \end{array} \\
   \begin{array}{*{20}c}
    1 \\ 2 \\ \vdots \\ l_{max} \\
   \end{array}
   &

\left[ {\begin{array}{*{20}c}
   {C(e_0^{(1)} )} & {C(e_1^{(1)} )} &  \cdots  & {C(e_{l_T  - 1}^{(1)} )}  \\
   {C(e_0^{(2)} )} & {C(e_1^{(2)} )} &  \cdots  & {C(e_{l_T  - 1}^{(2)} )}  \\
    \vdots  &  \vdots  &  \ddots  &  \vdots   \\
   {C(e_0^{(l_{max} )} )} & {C(e_1^{(l_{max} )} )} &  \cdots  & {C(e_{l_T  - 1}^{(l_{max} )} )}  \\
\end{array}} \right]

   \\
\end{array}  & &
\end{eqnarray}
where $C(e_k^{(i)} )$ is the error pattern (type $i$/ position
$k$) reliability measure computed by the maximum \emph{a
posteriori} (MAP)-based error-pattern correlator shown in
Fig.~\ref{Fig:Correlator}. The bank of local correlators discussed
here was also employed in~\cite{Jih_Moon} for AWGN channels, and
in~\cite{Hakim07} for data-dependent noise environments. We now
discuss how to generate these local metrics. Let $r_k$ be the
channel detector input sequence $r_k = {c}_k * h_k + w_k$, where
${c}_k$ is the bipolar representation of the recorded codeword
sequence, $h_k$ is the partial response channel of length $l_h$,
and $w_k$ is zero-mean AWGN noise with variance $\sigma^2$. Also,
let ${q}_k = r_k - (\hat{c}_k
* h_k) = (c_k - \hat{c}_k)*h_k
+ w_k $ be the channel detector's output error sequence. If a
target error pattern sequence $e_k^{(i)}$
 occurs at positions from $k = j$ to $k = j+l_i-1$,
 then $q_k$ can be written as
 \begin{align}\label{eq:qk}
 \begin{split}
 q_k &= [\textbf{c}- \hat{\textbf{c}}^{(i)}]_j^{j+l_i-1}*h_k + w_k\\
     &= [\textbf{e}^{(i)}]_j^{j+l_i-1}*h_k + w_k \\
     &= [\textbf{s}^{(i)}]_j^{j+l_i^h} + w_k
 \end{split}
 \end{align}
where $s_k^{(i)}$ is the channel response of the error sequence,
and is given by $s_k^{(i)} = e_k^{(i)}*h_k$, and $l_i^h = l_i +
l_h -2$. Note that we define the start of the tensor symbol at
$j=0$. So, if $j<0$, then the error pattern starting position is
in a preceding tensor symbol.

The reliability for each error pattern with starting position,
$j$, can be computed by the local \emph{a posteriori}
probabilities (ignoring tensor symbol boundaries for now):
\begin{eqnarray}\label{eq:postp}
\nonumber \Pr \left( {\left. {[\textbf{e}^{(i)}]_j^{j+l_i-1} }
\right|[\textbf{r}]_j^{j + l_i^h } ,[{\bf{\hat c}}]_{j - l_h  +
1}^{j + l_i^h } } \right) & & \\
 = \Pr \left( {\left. {[\textbf{s}^{(i)}]_j^{j + l_i^h } }
\right|[\textbf{q}]_j^{j + l_i^h } ,[{\bf{\hat c}}]_{j - l_h  +
1}^{j + l_i^h } } \right). & &
\end{eqnarray}
The most likely assumed error type/position pair in a tensor
symbol maximizes the \emph{a posteriori} probability ratio of its
reliability to the reliability of the most probable error event
(the competing event in this case would be the ML word itself,
with no error occurrence assumed at the output of Viterbi
detection). Hence, utilizing (\ref{eq:postp}) and Bayes rule, the
ratio to maximize becomes
\begin{eqnarray}\label{eq:ratio}
\displaystyle
 \begin{array}{*{1}l} \displaystyle
 \frac{{\Pr \left( {\left. {\textbf{e}_j^{(i)} } \right|{\bf{\hat
c}},[\textbf{q}]_j^{j + l_i^h }} \right)}}{{\Pr \left( {\left.
{\left[ {{\rm{\textbf{ML word}}}} \right]_j^{j+l_i-1} }
\right|{\bf{\hat c}},[\textbf{q}]_j^{j + l_i^h }} \right)}} = \\
\displaystyle \frac{{\Pr \left( {\left. {[{\bf{q}}]_j^{j + l_i^h }
} \right|[{\bf{\hat c}}]_{j - l_h  + 1}^{j + l_i^h }
,[{\bf{s}}^{(i)} ]_j^{j + l_i^h } } \right)\Pr \left(
{[{\bf{s}}^{(i)} ]_j^{j + l_i^h } } \right)}}{{\Pr \left( {\left.
{[{\bf{q}}]_j^{j + l_i^h } } \right|[{\bf{\hat c}}]_{j - l_h  +
1}^{j + l_i^h } ,[{\bf{\tilde s}}^{(i)} ]_j^{j + l_i^h } }
\right)\Pr \left( {[{\bf{\tilde s}}^{(i)} ]_j^{j + l_i^h } }
\right)}}
 \end{array}
\end{eqnarray}
where $[{\bf{\tilde s}}^{(i)} ]_j^{j + l_i^h}$ is the ML word's
noiseless channel response. Given the noise model,
$[{\bf{q}}]_j^{j + l_i^h }$ is a sequence of independent Gaussian
random variables with variance $\sigma^2$. Therefore, maximizing
(\ref{eq:ratio}) can be shown to be equivalent to maximizing the
log-likelihood local measure~\cite{Jih_Moon}:
\begin{equation}\label{eq:nobound}
C(e_j^{(i)}) = \sum\limits_{k = j}^{j + l_i^h } \frac{1}{{2\sigma
^2 }}\left( {q_k ^2  - (q_k  - s_k^{(i)} )^2 } \right) - \log
\frac{{\Pr ([{\bf{\tilde s}}^{(i)} ]_j^{j + l_i^h } )}}{{\Pr
([{\bf{s}}^{(i)} ]_j^{j + l_i^h } )}}
\end{equation}
where the \textit{a priori} bias in (\ref{eq:nobound}) is
evaluated as:
\begin{equation}\label{aprioribias}
\log \frac{{\Pr([{\bf{\tilde s}}^{(i)} ]_j^{j + l_i^h
})}}{{\Pr([{\bf{s}}^{(i)} ]_j^{j + l_i^h })}} = \sum\limits_{k =
j,\hat c_k =  + 1}^{j + l_i^h } {\lambda _k }  - \sum\limits_{k =
j,\hat c_k =  - 1}^{j + l_i^h } {\lambda _k }
\end{equation}
where $\lambda_k$ is the \textit{a priori} LLR of the error-event
bit at position $k$ as received from the outer soft decoder, and
we are assuming here that error event sequences do not include $0$
bits, i.e., the ML sequence and error sequence do not agree for
the entire duration of the error event. Equation
(\ref{eq:nobound}) represents the ``local" error-pattern
correlator output in the sense that it essentially describes the
correlator operation between $q_k$ and the channel output version
of the dominant error pattern $e_j^{(i)}$ within the local region
$[j, j + l_i^h]$. However, equation (\ref{eq:nobound}) ignores
that errors can span tensor symbol boundaries when $j<0$ or $j+
l_i - 1 > l_T-1$. For instance, an error in the first bit of the
tensor symbol can result from a single error event in that bit, a
double error event in the last bit of the preceding tensor symbol,
a triple error event occurring two bits into the previous symbol,
and so on. Hence, the probability of an error in the first bit is
the sum of all these parent error event probabilities. Moreover,
this can be easily generalized to boundary errors extending beyond
the first bit. In a similar manner, an error in the last bit of a
tensor symbol can result from a single error event in that bit, a
double error event starting in that bit and continuing into the
next tensor symbol, a triple error starting at the last bit and
continuing into the next tensor symbol, and so on. Again, the
probability of an error event in that bit is the sum of the
probabilities of all these parent events. Moreover, we have to
nullify the probability of the parent error events in the modified
reliability matrix since they are already accounted for in the
last bit's reliability calculation. Furthermore, this can also be
generalized to error events starting earlier than the last bit and
extending into the next tensor symbol. In summary, to calculate a
modified metric relevant to the current tensor symbol, we utilize
the following procedure:
\begin{itemize}
    \item $\forall i$ at $j=0$, modify $C(e_0^{(i)})$
    $$\tilde C(e_0^{(i)}) = \mathop {max^*}\limits_{k = i}^{l_{max}
} \left( {C(e_{ - k + i}^{(k)} )} \right),$$ independently for
each $i$, where $l_{max}$ is the maximum length of a targeted
error pattern.
    \item Starting at $i=1$ and $j=l_T-1$, do:
    \renewcommand{\labelenumi}{(\roman{enumi})}
    \begin{enumerate}
        \item $\tilde C(e_j^{(i)} ) = \mathop {max ^* }\limits_{k = i}^{l_{max}
} \left( {C(e_j^{(k)} )} \right)$.
        \item $\forall k > i$, set $\tilde C(e_j^{(k)} ) =  - \infty$.
        \item Set $i=i+1$, $j=j-1$.
        \item If $i < l_{max}$ go back to (i).
    \end{enumerate}
\end{itemize}
We assume here that dominant error events span only two tensor
symbols at a time and that they do not include error free gaps,
which is certainly true for the case study of this paper.
Following this procedure we obtain the modified reliability matrix
${\bf{\tilde{C}}}\left( E \right)$.
\subsubsection{Signature mlLLR Calculation}
For each tensor symbol $i$, utilizing ${\bf{\tilde{C}}}\left( E
\right)$, we need to find the p.m.f. or the log domain mlLLR of
its signature $Sg_i \in GF(2^{p_{epcc}})$, for EPCC with
$p_{epcc}$ parity bits. To limit the computational complexity of
this calculation, we construct a signature only from the dominant
errors and a subset of their multiple occurrences. Denote
$\hat{Pr}(Sg_i=\alpha^{j-1})$ as the running estimate of the
p.m.f. at $\alpha^{j-1}$, and
$\hat{\gamma}(Sg_i=\alpha^{j-1})=\log(\hat{Pr}(Sg_i=\alpha^{j-1}))$
as the running estimate of mlLLR. Denote a one dimensional index
of ${\bf{\tilde{C}}}\left( E \right)$ as $ p^{rc} = (p_c \times
l_{max}) + p_r $ corresponding to the $p_r$\textit{-th} row and
$p_c$\textit{-th} column of ${\bf{\tilde{C}}}\left( E \right)$ and
error $E(p^{rc})$. We choose the dominant list as the $L$ patterns
with the largest corresponding elements of ${\bf{\tilde{C}}}\left(
E \right)$ having indexes $\{p_i^{rc}\}^{i=L}_{i=1}$. Based on
this list, we developed the following procedure to compute
$\hat{\gamma}(Sg_i=\alpha^{j-1})$:
\begin{itemize}
    \item Step $1$ (Single occurrences):
     \begin{eqnarray}\label{eq:m1LLR1}
  \nonumber \begin{array}{*{10}c}
  \hat{\gamma}(Sg_i=\alpha^{j-1})= \mathop {max^* }\limits_{k = p_1^{rc}}^{p_L^{rc}}\left( {\tilde{C}\left(k\right)}
  \right), & & & & & &
  \end{array} & & \\ \nonumber
   \forall k : \mathop {G_f}\limits_{q_{epcc}}(H_{epcc}\times [\hat{c}_{i \times l_T}^{(i+1) \times l_T-1} \oplus E(k)]^t) =
   \alpha^{j-1} & & \\
\end{eqnarray}
where $q_{epcc}=2^{p_{epcc}}$, and $G_f(.)$ is an operator that
maps $p_{epcc}$-bit vectors into $GF(q_{epcc})$ symbols.
    \item Step $2$ (Double occurrences):
    \begin{eqnarray}\label{eq:m1LLR2}
 \nonumber \begin{array}{*{13}c}
 \hat{\gamma}(Sg_i=\alpha^{j-1}) = & & & & & & & & & & & &
  \end{array} \\ \nonumber
 max^*\left(
\begin{array}{*{1}c}
\hat{\gamma}\left(Sg_i=\alpha^{j-1}\right) \\
\left\{
{\tilde{C}\left(k\right) + \tilde{C}\left(m\right)}\right\}_{k =
p_1^{rc}, m = p_1^{rc}}^{p_L^{rc},p_L^{rc}}
\end{array}
\right),\\
 \nonumber  \forall \{k, m\} : \mathop \mathbb{D}\limits_{k \neq m}( E(k), E(m)) > E_{free}, \\
\nonumber  \mathop {G_f}\limits_{q_{epcc}}(H_{epcc}\times
[\hat{c}_{i \times l_T}^{(i+1) \times l_T-1} \oplus E(k) \oplus
E(m)]^t) = \alpha^{j-1} \\
\end{eqnarray}
where $\mathbb{D}$ is the error free distance between the two
errors, $E_{free}=l_h-1$ is the error free distance of the channel
beyond which the errors are independent.
    \item ...
    \item Step $M$ ($M$ occurrences):
    \begin{eqnarray}\label{eq:m1LLRM}
\nonumber \begin{array}{*{13}c} \hat{\gamma}(Sg_i=\alpha^{j-1}) =
& & & & & & & & & & & &
\end{array} \\ \nonumber max^*\left(
\begin{array}{*{1}c}
\hat{\gamma}\left(Sg_i=\alpha^{j-1}\right) \\ \left\{
{\sum_{\xi=1}^{M}\tilde{C}\left(q_{\xi}\right)}\right\}_{q_{\xi} =
p_1^{rc}, \xi=1,...,M}^{p_L^{rc}}
\end{array}
\right), \\
\nonumber \forall \{q_1, q_2, ..., q_M \} :  \mathop \mathbb{D}\limits_{s, t, s \neq t}( E(q_s), E(q_t)) > E_{free}, \\
\nonumber \mathop {G_f}\limits_{q_{epcc}}(H_{epcc}\times [\mathop
\bigoplus \limits_{\xi=1}^{\xi=M} E(q_{\xi}) \oplus \hat{c}_{i
\times l_T}^{(i+1) \times l_T-1}]^t) = \alpha^{j-1} \\
\end{eqnarray}

    \item Step $M+1$ (ML-signature reliability; computed so that the resulting
    signature p.m.f. sums to $1$):
    \begin{eqnarray}\label{eq:preSgML}
    \nonumber \begin{array}{*{16}l}
    \breve{\gamma}\left( Sg_i=\alpha^{\beta_{ML}} \right) = & & & & & & & & & & & &
    \end{array}
    \\
    -max^* \left(\begin{array}{*{1}c}0\\
    \mathop {max^* } \limits_{j = -\infty,j \neq \beta_{ML}+1}^{q_{epcc}-1} \hat{\gamma}
\left( Sg_i=\alpha^{j-1} \right)
\end{array}
\right)
     \end{eqnarray}
    \begin{eqnarray}\label{eq:SgML}
    \hat{\gamma} \left( Sg_i=\alpha^{\beta_{ML}} \right)= max^* \left(
    \begin{array}{*{1}c}
    \breve{\gamma} \left(Sg_i=\alpha^{\beta_{ML}} \right) \\ \hat{\gamma} \left( Sg_i=\alpha^{\beta_{ML}} \right)
\end{array}
\right)
\end{eqnarray}
    \begin{eqnarray}\label{eq:pmf1}
     \nonumber \hat{\gamma}(Sg_i=\alpha^{j-1})=\hat{\gamma}(Sg_i=\alpha^{j-1})+\hat{\gamma}(Sg_i=\alpha^{\beta_{ML}}),\\
     \nonumber j=-\infty, 1, ..., q_{epcc}-1; j \neq \beta_{ML}+1.
     \\
     \end{eqnarray}

    \item Step $M+2$ (Normalization):
    \begin{equation}\label{eq:norm}
     \gamma(Sg_i=\alpha^{j-1})=\hat{\gamma}(Sg_i=\alpha^{j-1})-\hat{\gamma}(Sg_i=\alpha^{-\infty}).
     \end{equation}

\end{itemize}

In steps $1$ through $M$, to calculate the log-likelihood of
signature $i$ assuming value $\alpha^{j-1}$, we sum the
probabilities of all presumed single and multiple errors in the ML
word whose signatures equal $\alpha^{j-1}$. This is equivalent to
performing the $max^*$ operation in the log domain on error
reliabilities dictated by ${\bf{\tilde{C}}}\left( E \right)$.
However, to limit the complexity of this stage, we only use a
truncated set of possible error combinations, in all steps from
$1$ to $M$. Also, for signature values that do not correspond to
any of the combinations, we set their reliability to $-\infty$, or
more precisely, a reasonably large negative value in practical
decoder implementation. Since there are many such signature
values, the corresponding constructed p.m.f. will be sparse.

In step $M+1$, the likelihood of the ML signature value is
computed so that the p.m.f. of the tensor symbol signature sums to
$1$. In this step, the $max^*$ operation in (\ref{eq:SgML}) is a
reflection of the fact that in previous steps, $1$ through $M$,
some multiple error occurrences have the same signature as the ML
tensor symbol value. So, we have to account for such error
instances in the running estimate of the ML signature reliability.
These events correspond to cases where error events are not
detectable by $H_{epcc}$, i.e., they belong to the null space of
$H_{epcc}$. In step $M+2$, the mlLLR of the tensor symbol is
centered around $\hat{\gamma}(Sg_i=0)$ to prevent the $q$LDPC SPA
messages from saturating after a few BP iterations.

\subsubsection{$q$-ary LDPC Decoding}
Now, the sequence of signature mlLLRs is passed as multi-level
channel observations to the $q$LDPC decoder. We choose to
implement the log-domain $q$-ary fast Fourier transform-based SPA
(FFT-SPA) decoder in~\cite{Hongxin03} for this purpose. The choice
of log-domain decoding is essential, since if we use the signature
p.m.f. as input, the SPA would run into numerical instability
resulting from the sparse p.m.f. generated by the preceding stage.

The LDPC output \textit{posteriori} mlLLRs correspond to the
signatures of tensor symbols, rather than the syndromes of errors
expected by EPCC decoding. Similar to the decoder of T-EPCC-RS,
error-syndrome $Syn_i^e$ is the finite field sum of the LDPC's
input channel observation of signature $i$, $Sg_i^{ch}$, and
output \textit{posteriori} signature reliability, $Sg_i^p$.
Moreover, the addition of hard signatures corresponds to the
convolution of their p.m.f.'s, and this convolution in probability
domain corresponds to the following operation in log-domain:
\begin{eqnarray}\label{eq:Conv}
\nonumber \begin{array}{*{13}c}
\hat{\gamma}(Syn_i^e=\alpha^{\beta_{e}}) = & & & & & & & & & & & &
\end{array}
\\ \nonumber
{max^*}\left(
\begin{array}{*{1}c}
\hat{\gamma}(Syn_i^e=\alpha^{\beta_{e}})
\\ \gamma(Sg_i^{ch}=\alpha^{\beta_{ch}})+\gamma(Sg_i^p=\alpha^{\beta_{p}})
\end{array}
 \right),
\\ \nonumber
\forall \{\beta_{ch},\beta_{p}\}: \alpha^{\beta_{e}} =
\alpha^{\beta_{ch}} \mathop \oplus \limits_{GF(q_{epcc})}
\alpha^{\beta_{p}},\\ \nonumber \beta_{ch} = -\infty, 0, ...,
q_{epcc}-2;\\ \nonumber
 \beta_{p} = -\infty, 0,...,q_{epcc}-2.\\
\end{eqnarray}
The error-syndrome mlLLR is later normalized, similar to LDPC BP
mlLLR message normalization, according to:
\begin{eqnarray}\label{eq:Norm}
\nonumber
\gamma(Syn_i^e=\alpha^{\beta_{e}})=\hat{\gamma}(Syn_i^e=\alpha^{\beta_{e}})-\hat{\gamma}(Syn_i^e=\alpha^{-\infty}),
\\
\forall \beta_{e} = -\infty, 0, ..., q_{epcc}-2.
\end{eqnarray}
\subsubsection{EPCC Decoding}
An error-syndrome will decode to many possible error events due to
the low minimum distance of single-error correcting EPCC. However,
EPCC relies on local channel side information to implement a
list-decoding-like procedure that enhances its multiple error
correction capability. Moreover, the short codeword length of EPCC
reduces the probability of such multiple error occurrences
considerably. To minimize power consumption, EPCC is turned on for
a tensor symbol $i$ only if the most likely value of the
error-syndrome mlLLR is nonzero, i.e., $\mathop {\arg \max
}\limits_{\alpha^{\beta} \in GF(q_{epcc} )} \gamma (Syn_i^e  =
\alpha ^{\beta} ) \ne \alpha ^{ - \infty }$, indicating that a
resolvable error has occurred. After this, a few syndrome values,
$3$ in our case, most likely according to the mlLLR, are decoded
in parallel. For each of these syndromes, the list decoding
algorithm goes as~\cite{Jih_Moon,Hakim07}:
\begin{itemize}
    \item A test error word list is generated by inserting the most probable combination of local error patterns into the ML tensor
    symbol.
    \item An array of parallel EPCC single-pattern correcting decoders decodes the
    test words to produce a list of valid codewords that satisfy the current error-syndrome.
    \item The probability of a candidate codeword is computed as the sum of likelihoods of its parent
    test-word and the error pattern separating the two.
    \item Each candidate codeword probability is biased by the
    likelihood of the error-syndrome it is supposed to satisfy.
\end{itemize}
In addition, when generating test words, we only combine
independent error patterns that are separated by the error free
distance of the ISI channel.
\subsubsection{Soft Bit-level Feedback LLR Calculation}
The list of candidate codewords and probabilities are used to
generate bit level-probabilities in a similar manner to
~\cite{Pyndiah98,Hakim07}. The conversion of word-level
reliability into bit-level reliability for a given bit position
can be done by grouping the candidate codewords into two groups,
according to the binary value of the hard decision bit in that bit
position, and then performing group-wise summing of the word-level
probabilities. Three scenarios are possible for this calculation:

\renewcommand{\labelenumi}{(\roman{enumi})}
\begin{enumerate}
\item The candidate codewords do not all agree on the bit decision
for location $k$; then, given the list of codewords and their
accompanying \emph{a posteriori} probabilities, the reliability
$\lambda_k$ of the coded bit $c_k$ is evaluated as
\begin{eqnarray}\label{eq:3}
  \lambda_k=\log\frac{\sum_{\mathbf{c} \in \textbf{S}_{k}^{+}}Pr(\mathbf{c}|\mathbf{\hat{c}},\mathbf{r})}{\sum_{\mathbf{c} \in
  \textbf{S}_{k}^{-}}Pr(\mathbf{c}|\mathbf{\hat{c}},\mathbf{r})}
\end{eqnarray}
where $\textbf{S}_{k}^{+}$ is the set of candidate codewords where
$c_k=+1$, and $\textbf{S}_{k}^{-}$ is the set of candidate
codewords where $c_k=-1$
    \item  Although rare for such short codeword
lengths, in the event that all codewords do agree on the decision
for $c_k$, a method inspired by~\cite{Pyndiah98} is adopted for
generating soft information as follows
\begin{eqnarray}\label{eq:4}
  \lambda_k=\beta^{iter} \times \lambda_{max} \times \hat{d}_{k}
\end{eqnarray}
where $\hat{d}_k$ is the bipolar representation of the agreed-upon
decision, $\lambda_{max}$ is a preset value for the maximum
reliability at convergence of turbo performance, and the
multiplier $\beta^{iter} < 1$ is a scaling factor.
$\beta^{iter}\ll 1$ in the first global iterations and is
increased to $1$ as more global iterations are performed and the
confidence in bit decisions improved. Thus, this back-off control
process reduces the risk of error propagation.
    \item The heuristic scaling in (\ref{eq:4}) is again useful when
EPCC is turned off for a tensor symbol, in case the most likely
error-syndrome being $0$. Then, the base hard value of the tensor
symbol corresponds to the most likely error event found as a side
product in stage $2$ of the T-EPCC-$q$LDPC decoder.
\end{enumerate}

\subsection{Stopping Criterion for T-EPCC-$q$LDPC and RS Erasure
Decoding} Due to the ambiguity in mapping tensor symbols to
signatures and syndromes to errors in stages $2$ and $4$ of the
decoder, respectively, the possibility of non-targeted error
patterns, or errors that have zero error-syndromes that are
transparent to $H_{epcc}$, a second line of defense is essential
to take care of undetected errors. Therefore, an outer RS code of
small correction power $t_{out}$ is concatenated to T-EPCC-$q$LDPC
to take care of the imperfections of the component EPCC. Several
concurrent functions are offered by this code, including:
\begin{itemize}
    \item \textit{Stopping Flag}: If the RS syndrome is zero, then, global iterations are halted and decisions are released.
    \item \textit{Outer ECC}: Attempt to correct residual errors at the output of EPCC after each global iteration.
    \item \textit{Erasure Decoding}: If the RS syndrome is nonzero, then, for
    those tensor symbols that EPCC was turned on, declare
    their bits as erasures. Next, find the corresponding RS symbol erasures, and attempt RS erasure decoding which is capable of
    correcting up to $2\times t_{out}$ such erasures. In this case, T-EPCC acts as an error locating code.
\end{itemize}

\section{Simulation Results and Discussion} \label{secIX}
We compare three coding systems based on LDPC: conventional binary
LDPC, $q$-ary LDPC, and T-EPCC-$q$LDPC, where all the component
LDPC codes are regular and constructed by PEG with a QC
constraint. We study their sector error rate (SER) performance on
the ideal equalized partial response target $1+0.85D$ corrupted by
AWGN, and with coding rate penalty $10\log_{10}(1/R)$. The nominal
systems run at a coding rate of $0.9$. The minimum SNR required to
achieve reliable recording at this rate is $3.9$ dB, estimated by
following the same approach as in~\cite{Arnold01}.

\subsection{Single-level BLDPC \& $q$LDPC Simulation Results}
\begin{figure}[h]
\centering
\includegraphics*[width=3.5in]{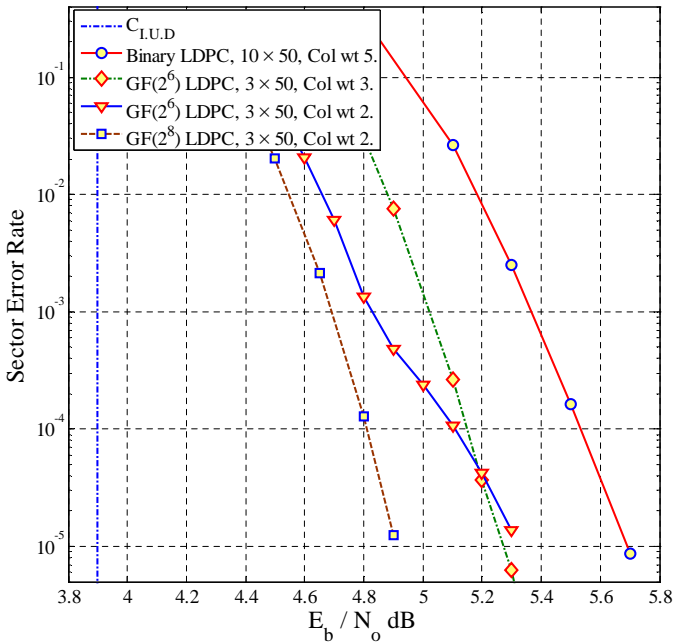}
\caption{Comparing SER of: $10 \times 50$ iterations of binary
LDPC, $3 \times 50$ iterations of $GF(2^8)$-LDPC of column weight
$2$, and $3 \times 50$ iterations of $GF(2^6)$-LDPC of column
weights $2$ and $3$. Minimum SNR to achieve reliable recording at
coding rate $0.9$ is $3.9$ dB for $1+0.85D$.}
\label{Fig:LDPCalone}
\end{figure}

In Fig.~\ref{Fig:LDPCalone}, we compare SER of the following LDPC
codes, each constructed by PEG with a QC constraint:
\begin{itemize}
    \item A $(4550,4095)$ $GF(2)$-LDPC, of column weight $5$, and circulant size
    $91$ bits. The channel detector is a $2$ state binary BCJR.
    \item A $(570,510)$ $GF(2^8)$-LDPC, of codeword length $4560$ bits, column weight $2$, and circulant size of $15$
    symbols. The channel detector is a symbol-BCJR with $256$ branches emanating from each of $2$ states.
    \item A $(760,684)$ $GF(2^6)$-LDPC, of codeword length $4560$ bits, column weight $2$, and circulant size of $19$ symbols.
    The channel detector is a symbol-BCJR with $64$ branches emanating from each of $2$ states.
    \item A $(775,700)$ $GF(2^6)$-LDPC, of codeword length $4650$ bits, column weight $3$, and circulant size of
    $25$ symbols. The channel detector is a symbol-BCJR with $64$ branches emanating from each of $2$ states.
\end{itemize}
For the binary LDPC turbo equalizer, we run a maximum of
$10\times50$ iterations, $10$ global, and $50$ LDPC BP iterations.
For the $q$-ary turbo equalizers, on the other hand, we run a
maximum of $3\times50$ iterations. A column weight of $2$ gives
the best waterfall performance of $q$-ary LDPC. However,
$GF(2^6)$-LDPC exhibits an error floor as early as at SER
$6\times10^{-4}$, whereas a higher order field of $GF(2^8)$ does
not show such a tendency down to $1\times10^{-5}$. Nevertheless,
the prohibitive complexity of $GF(2^8)$ symbol-BCJR makes
$GF(2^6)$-LDPC a more attractive choice. Still, we need to
sacrifice $GF(2^6)$-LDPC's waterfall performance gains to
guarantee a lower error floor. For that purpose, we move to a
column weight $3$ $GF(2^6)$-LDPC that is $1.37$ dB away at
$1\times10^{-5}$ from the independent uniformly-distributed
capacity $C_{I.U.D.}$ of the channel~\cite{Arnold01}, and $0.37$
dB away from $GF(2^8)$-LDPC a the same SER. In this simulation
study, we have observed that while binary LDPC can gain up to
$0.4$ dB through $10$ channel iterations before gain saturates,
$GF(2^8)$-LDPC and $GF(2^6)$-LDPC achieve very little iterative
gain by going back to the channel, between $0.09$ to $0.12$ dB
through $3$ channel iterations. One way to explain this
phenomenon, is that symbol-level LDPC decoding divides the bit
stream into LDPC symbols that capture the error events introduced
by the channel detector, rendering the binary inter-symbol
interference limited channel into a memoryless multi-level AWGN
limited channel. Nonetheless, error events spanning symbol
boundaries reintroduce correlations between LDPC symbols that are
broken only by going back to the channel. In other words, if it
was not due to such boundary effects, a $q$-ary LDPC equalizer
would not exhibit any iterative turbo gain whatsoever.
Nonetheless, full-blown symbol BCJR is still too complex to
justify salvaging the small iterative gain by performing extra
channel iterations~\cite{Chang4k08}. This is where error event
matched decoding comes into the picture, which leads us to the
results of the next section.
\subsection{T-EPCC-$q$LDPC Simulation Results}
We first construct two T-EPCC-$q$LDPC codes of rate $0.9$, the
same rate as the competing single-level $q$LDPC. These TPPC's are
based on EPCC $(12,6)$ of example 1. The codes constructed are:
\begin{itemize}
\item TPPC-A: A $1/2$ KB sector, binary $(4680,4212)$ TPPC, of
rate $0.9$, and $468$ parity bits, based on a component
$(390,312)$ PEG-optimized QC $GF(2^6)$-LDPC, of rate $0.8$, column
weight $3$, and circulant size $26$.

\item TPPC-B: A $1$ KB sector, binary $(9360,8424)$ TPPC, of rate
$0.9$, and $936$ parity bits, based on a component $(780,624)$
PEG-optimized QC $GF(2^6)$-LDPC, of rate $0.8$, column weight $3$,
and circulant size $52$.
\end{itemize}

\begin{figure}[h]
\centering
\includegraphics*[width=3.5in]{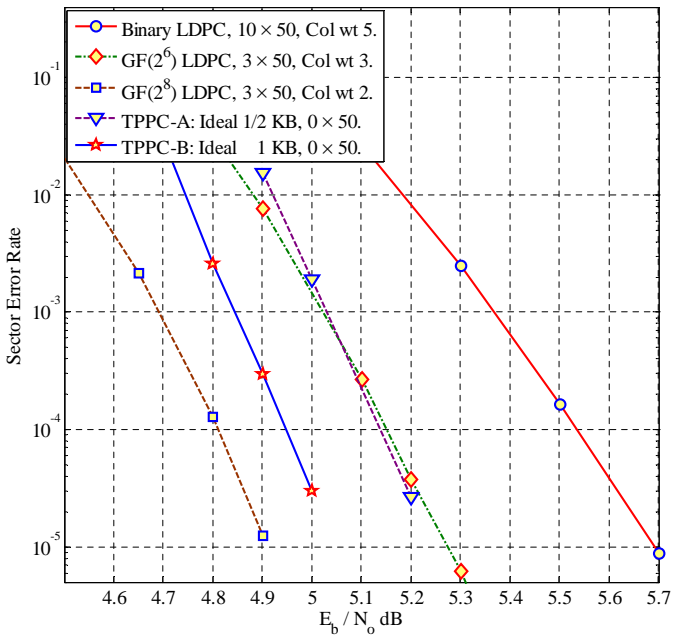}
\caption{Comparing SER of: $10 \times 50$ iterations of binary
LDPC, $3 \times 50$ iterations of $GF(2^8)$-LDPC of column weight
$2$, $3 \times 50$ iterations of $GF(2^6)$-LDPC of column weight
$3$, and $0 \times 50$ iterations of ideal $1/2$ KB and $1$ KB
T-EPCC-$GF(2^6)$LDPC based on column weight $3$ LDPC.}
\label{Fig:idealTPPC}
\end{figure}

First, we study the SER of T-EPCC-$q$LDPC just up to the component
$GF(2^6)$ LDPC decoder, and only at the first channel pass. This
SER is function of the Viterbi symbol error rate, and the accuracy
of generating signature mlLLRs, in addition to the component LDPC
employed. This SER represents the best that the TPPC code can do,
under the assumption of perfect component EPCC, i.e., as long as
$q$LDPC generates a clean codeword of signature-symbols, then EPCC
generates a clean codeword of data-symbols.
Fig.~\ref{Fig:idealTPPC} shows the ideal SER of these two TPPC
codes, assuming perfect EPCC, compared to single-level
$GF(2^6)$-LDPC and $GF(2^8)$-LDPC. Ideal $1/2$ KB TPPC has about
the same SER as single level $GF(2^6)$-LDPC at $3\times10^{-5}$
SER. In $1/2$ KB TPPC, the component $GF(2^6)$-LDPC has half the
codeword length of the single level counterpart, saving $50\%$ of
the decoder complexity, while delivering similar SER performance.
The TPPC component $q$LDPC faces a harsher channel than
single-level $q$LDPC, because the symbol error probability of
$6$-bit data symbols is strictly less than the symbol error
probability of $6$-bit signature symbols, where signature symbols
are compressed down from $12$-bit data symbols. Also, the shorter
codeword length of component $q$LDPC hurts its minimum distance.
Still, these impairments are effectively compensated for by an
$11.4\%$ increase in the redundancy of the TPPC component LDPC. On
the other hand, if we match the codeword length of TPPC's
component LDPC to single-level LDPC, as part of constructing $1$
KB TPPC, then, $1$ KB TPPC will have similar decoder complexity to
$1/2$ KB single-level LDPC with about $0.2$ dB SNR advantage for
$1$ KB ideal TPPC at $3\times10^{-5}$ SER.

Due to the imperfections of EPCC design, including mis-correction
due to one-to-many syndrome to error position mapping, and
undetected errors due to EPCC's small minimum distance, achieving
the ideal performance in Fig.~\ref{Fig:idealTPPC} is not possible
in one channel pass. In addition, an outer code is necessary to
protect against undetected errors and provide a stopping flag for
the iterative decoder. Hence, one can think of an implementation
of the full T-EPCC-$GF(2^6)$LDPC decoder that includes an outer
$t=6$ $(421,409)$ RS for the $1/2$ KB case, and an outer $t=12$
$(842,818)$ RS for the $1$ KB case, so as to protect against EPCC
residual errors. These outer RS codes are defined on $GF(2^{10})$
and have rate $0.972$. However, this concatenation setup will run
at a lower code rate of $0.875$, which can incur an SNR
degradation larger than $0.25$ dB for a noise environment
characterized by the rate penalty $10\log_{10}(1/R^{\delta}),\
\delta \geq 2$. In a more thoughtful approach, one can preserve
the nominal code rate of $0.9$ and redistribute the redundancy
between the inner TPPC and outer RS to achieve an improved
tradeoff between miscorrection probability and the inner TPPC's
component LDPC code strength. In that spirit, we construct the
following concatenated codes:
\begin{itemize}
\item TPPC-C: A $1/2$ KB sector, binary $(4560,4218)$ TPPC, of
rate $0.925$, and $342$ parity bits, based on a component
$(380,323)$ PEG-optimized QC $GF(2^6)$-LDPC, of rate $0.85$,
column weight $3$, and circulant size $19$. An outer $t=6$
$(422,410)$ RS code of rate $0.972$ is included, resulting in a
total system rate of $0.9$.

\item TPPC-D: A $1$ KB sector, binary $(9120,8436)$ TPPC, of rate
$0.925$, and $684$ parity bits, based on a component $(760,646)$
PEG-optimized QC $GF(2^6)$-LDPC, of rate $0.85$, column weight
$3$, and circulant size $38$. An outer $t=12$ $(844,820)$ RS code
of rate $0.972$ is included, resulting in a total system rate of
$0.9$.
\end{itemize}

The control mechanism of iterative decoding for these codes is as
follows: if EPCC results in less than $6$ RS symbol errors for the
$1/2$ KB design or less than $12$ for the $1$ KB design, or if
EPCC generates more errors than this, but declares less than $12$
erasures for $1/2$ KB or $24$ erasures for $1$ KB, then, decoding
halts and decisions are released. Otherwise, one more channel
iteration is done by passing EPCC soft bit-level LLR's to Viterbi
detection and the bank of error-matched correlators.

\begin{figure}[h]
\centering
\includegraphics*[width=3.5in]{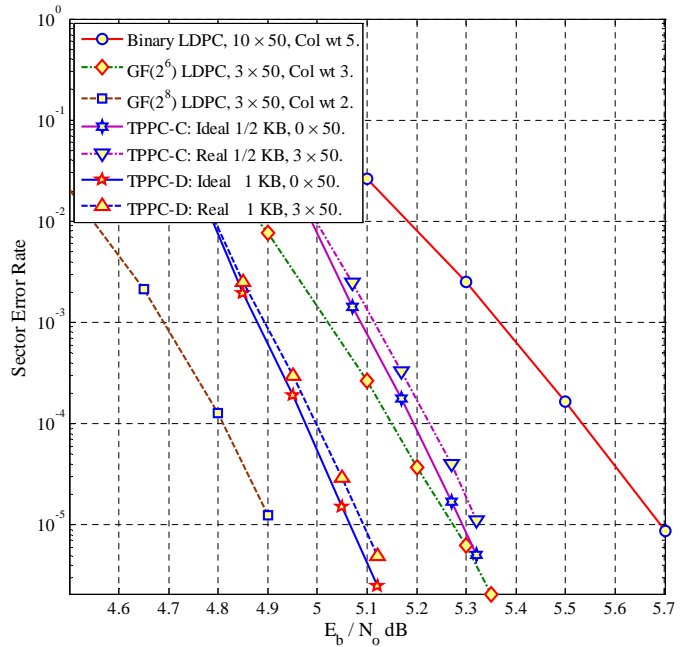}
\caption{Comparing SER in environment of rate penalty$10\log_{10}
1/R$: $10 \times 50$ iterations of binary LDPC, $3 \times 50$
iterations of $GF(2^8)$-LDPC of column weight $2$, $3 \times 50$
iterations of $GF(2^6)$-LDPC of column weight $3$, and $3 \times
50$ iterations of practical $1/2$ KB
T-EPCC-$GF(2^6)$LDPC+RS($t=6$), and $1$ KB
T-EPCC-$GF(2^6)$LDPC+RS($t=12$), both based on column weight $3$
LDPC.} \label{Fig:RealTPPC_R1}
\end{figure}

Simulation results in Fig.~\ref{Fig:RealTPPC_R1}, for a noise
environment of rate penalty $10\log_{10} 1/R$, demonstrate that
after $3$ channel iterations, the ideal and practical performances
of the new TPPC codes almost lock, while incurring minimal SNR
degradation. Also, $1/2$ KB TPPC saves $50\%$ of decoder
complexity while achieving the same SER performance as single
level LDPC for an additional SNR cost of $0.04$ dB at SER
$1\times10^{-5}$. Hence, TPPC-C represents a tradeoff between the
lower complexity of $GF(2)$-LDPC and performance advantage of
$GF(2^6)$-LDPC, whereas $1$ KB TPPC has the same decoding
complexity as single-level LDPC while furnishing $0.18$ dB gain at
$5\times10^{-6}$ SER. In terms of channel detector implementation
complexity, the complexity and latency of $GF(2^6)$-BCJR in the
single level code far exceeds the overall complexity of the
non-LDPC parts of two level T-EPCC-$GF(2^6)$LDPC including Viterbi
detection. At the same time, signature mlLLR generation, EPCC
decoding, and bit-LLR generation are all implemented tensor-symbol
by tensor-symbol, achieving full parallelism on the tensor-symbol
level. Furthermore, it is only when $q$LDPC finds a syndrome error
that EPCC decoding is turned on for each tensor symbol. To
eliminate redundant computations in the iterative decoder, branch
metric computation in Viterbi and (\ref{eq:nobound}) is only
required at the first pass. For all subsequent iterations,
however, only the \textit{a priori} bias is updated in the second
term of (\ref{eq:nobound}), and the branch update of
Viterbi~\cite{ChangCom08}.
One very important feature of the TPPC setup, that single-level
LDPC lacks, is its robustness to boundary error events. The
presence of a syndrome-constraint means that errors spanning
boundaries are broken by EPCC when attempting to independently
satisfy the adjacent tensor symbol syndromes, then, in the next
turbo iteration, adjacent tensor-symbols are decorrelated. This
mechanism enables TPPC to recover from these errors by iterative
decoding. However, for errors with a zero error-syndrome which go
undetected by EPCC, outer RS protection becomes handy.

Based on the fact that TPPC enables an increase in the redundancy
of its component LDPC, in addition to simulation results
demonstrating the utility of such lowered rate in combating the
harsher compressed channel, we conjecture that as the sector
length of both TPPC and single-level LDPC is driven to infinity,
TPPC will achieve strict error rate SNR gains. This is mainly
because of its surplus of redundancy compared to the single level
code at the same rate penalty, whereas channel conditions and EPCC
correction power do not change with replication of tensor symbols,
and the error rate performance of LDPC asymptotically approaches
the noise threshold in the limit of infinite codeword length.
Therefore, within a channel-capacity achieving argument, in the
limit of infinite codeword length, we take the view that TPPC will
bridge the gap to capacity further than any single level system
could. Moreover, the advantage of TPPC for larger sector sizes is
more timely than ever as the industry moves to the larger $4$ KB
sector format~\cite{Chang4k08}.

\section{Conclusions}\label{secXI}
In a tensor product setup, codes of short codeword length and low
rate can be combined into high rate codes of nice algebraic
properties. We showed that encoding of tensor product codes is
linear time if the component codes are linear time encodable. We
also demonstrated how the codeword length and rate of channel
matched EPCC can be substantially increased by combining with a
strong RS or LDPC of short codeword length. We also incorporated
an outer RS code of low correction power to clean out the residual
errors of T-EPCC-RS or T-EPCC-LDPC TPPCs. In conclusion, this work
established T-EPCC-$q$LDPC as a reasonable complexity approach to
introducing non-binary LDPC to the perpendicular recording read
channel architecture, paving the way to reliable higher recording
densities.

\section*{Acknowledgment}

The authors would like to thank the anonymous reviewers for their
constructive comments that helped enhance the technical quality
and presentation of this paper.

\vspace {-1.5in}
\begin{biography}[{\includegraphics[width=1in,height=1.25in,clip,keepaspectratio]{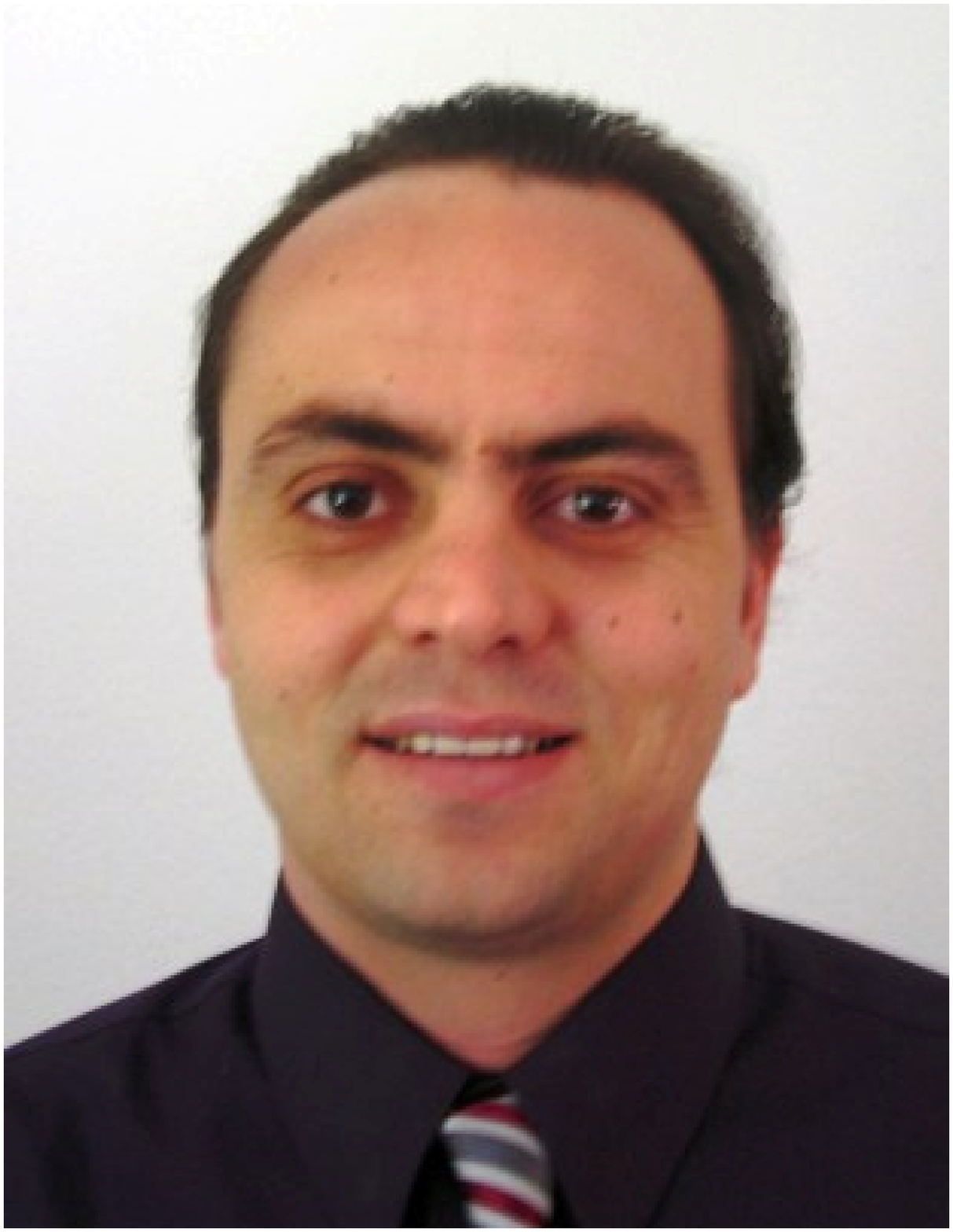}}]{Hakim Alhussien} received the B.S. and M.S. degrees in electrical
engineering with high honors from Jordan University of Science and
Technology (JUST), Irbid, in 2001 and 2003, respectively, and the
MSEE and Ph.D. degrees in electrical engineering from the
University of Minnesota Twin-Cities, Minneapolis, in 2008 and
2009, respectively. From 2003 to 2004, he was an Instructor with
the Department of Electrical Engineering at the University of
Yarmouk, Jordan. From 2004 to 2008, he held Research and Teaching
Assistant positions with the Department of Electrical and Computer
Engineering, the University of Minnesota. Since September 2008, he
has been a Systems Design Engineer with Link-A-Media Devices,
Santa Clara. His main research interest are in the applications of
coding, signal processing, and information theory to data storage
systems.
\end{biography}
\vspace {-1.5in}
\begin{biography}[{\includegraphics[width=1in,height=1.25in,clip,keepaspectratio]{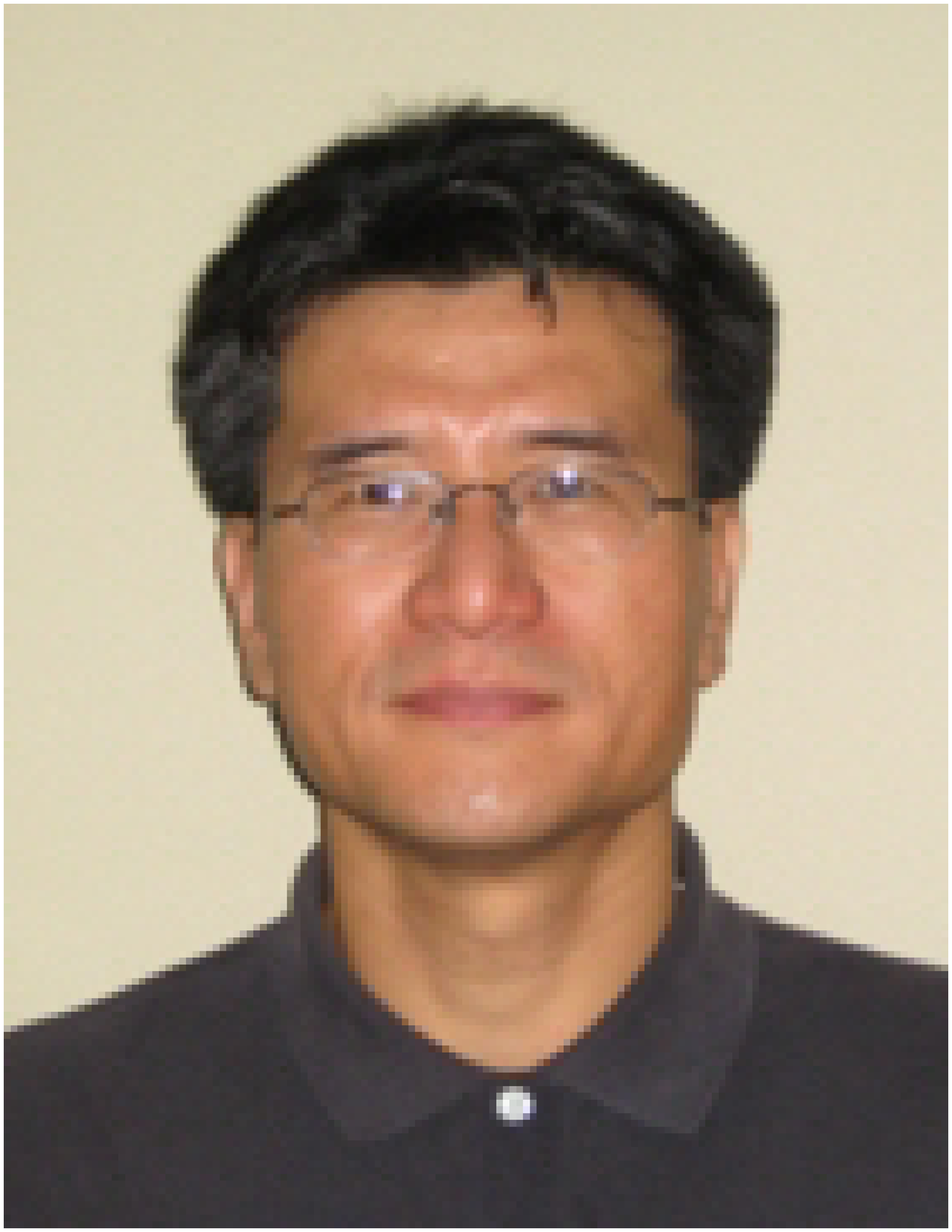}}]{Jaekyun Moon} is a Professor of Electrical Engineering at KAIST. Prof. Moon
received a BSEE degree with high honor from SUNY Stony Brook and
then M.S. and Ph.D. degrees in Electrical and Computer Engineering
at Carnegie Mellon University. From 1990 through early 2009, he
was with the faculty of the Department of Electrical and Computer
Engineering at the University of Minnesota, Twin Cities. Prof.
Moon's research interests are in the area of channel
characterization, signal processing and coding for data storage
and digital communication. He received the 1994-1996 McKnight
Land-Grant Professorship from the University of Minnesota. He also
received the IBM Faculty Development Awards as well as the IBM
Partnership Awards. He was awarded the National Storage Industry
Consortium (NSIC) Technical Achievement Award for the invention of
the maximum transition run (MTR) code, a widely-used
error-control/modulation code in commercial storage systems. He
served as Program Chair for the 1997 IEEE Magnetic Recording
Conference. He is also Past Chair of the Signal Processing for
Storage Technical Committee of the IEEE Communications Society. In
2001, he co-founded Bermai, Inc., a fabless semiconductor
start-up, and served as founding President and CTO. He served as a
guest Editor for the 2001 IEEE J-SAC issue on Signal Processing
for High Density Recording. He also served as an Editor for IEEE
Transactions on Magnetics in the area of signal processing and
coding for 2001-2006. He worked as consulting Chief Scientist at
DSPG, Inc. from 2004 to 2007. He also worked as Chief Technology
Officer at Link-A-Media Devices Corp. in 2008. He is an IEEE
Fellow.
\end{biography}


\begin{thebibliography}{11}

\bibitem{HaoDec07} Hao Zhong, Wei Xu, Ningde Xie, and Tong Zhang, ``Area-efficient min-sum
decoder design for high-rate quasi-cyclic low-density parity-check
codes in magnetic recording," \textit{IEEE Transactions on
Magnetics}, vol. 43, no. 12, pp. 4117-4122, Dec. 2007.

\bibitem{HaoMarch07} Hao Zhong, Tong Zhong, and Erich F. Haratsch, ``Quasi-cyclic LDPC codes
for the magnetic recording channel: Code design and VLSI
implementation," \textit{IEEE Transactions on Magnetics}, vol. 43,
no. 3, pp. 1118-1123, Mar. 2007.

\bibitem{Varnica03} N. Varnica, and A. Kavcic, ``Optimized low-density parity-check codes
for partial response channels," \textit{IEEE Communications
Letters}, vol. 7, no. 4, pp. 168-170, Apr. 2003.

\bibitem{Sanka03} S. Sankaranarayanan, B. Vasic, and E. M.Kurtas, ``Irregular
low-density parity-check codes: Construction and performance on
perpendicular magnetic recording channels," \textit{IEEE
Transactions on Magnetics}, vol. 39, no. 5, pp. 2567-2569, Sept.
2003.

\bibitem{Vasic04} B. Vasic, and O. Milenkovic, ``Combinatorial constructions of
low-density parity-check codes for iterative decoding,"
\textit{IEEE Transactions on Information Theory}, vol. 50, no. 6,
pp. 1156-1176, June 2004.

\bibitem{Liva08} G. Liva, G., W. E. Ryan, and M. Chiani, ``Quasi-cyclic generalized
LDPC codes with low error floors," \textit{IEEE Transactions on
Communications}, vol. 56, no. 1, pp. 49-57, Jan. 2008.

\bibitem{Yang04} M. Yang, W. E. Ryan, and Y. Li, ``Design of efficiently encodable
moderate-length high-rate irregular LDPC codes," \textit{IEEE
Transactions on Communications}, pp. 564-571, Apr. 2004.

\bibitem{Zongwang06} L. Zongwang, L. Chen, L. Zeng, S. Lin, and W. H. Fong, ``Efficient encoding of quasi-cyclic low-density parity-check
codes," \textit{IEEE Transactions on Communications}, vol. 54, no.
1, pp. 71-81, Jan. 2006.

\bibitem{Han06} Y. Han, and W. E. Ryan, ``Concatenating a structured
LDPC code and an RLL code to preserve soft-decoding, structure,
and burst correction," \textit{IEEE Trans. Magnetics}, pp.
2558-2560, Oct. 2006.

\bibitem{Kurkoski03} B. M. Kurkoski, P. H. Siegel, and J. K. Wolf, ``Joint message-passing
decoding of LDPC codes and partial-response channels,"
\textit{IEEE Transactions on Information Theory}, vol. 49, no. 8,
pp. 2076-2076, Aug. 2003.

\bibitem{Han08} Y. Han, and W. E. Ryan, ``LDPC decoder strategies for achieving low
error floors," \textit{Information Theory and Applications
Workshop, 2008}, pp. 277-286, Jan. 27 2008-Feb. 1 2008.

\bibitem{Cideciyan02}  R. D. Cideciyan, E. Eleftheriou, and T. Mittelholzer, ``Perpendicular and longitudinal recording: A signal-processing and coding perspective," \textit{IEEE Trans.
Magn.}, vol. 38, no. 4, pp. 1698-1704, Jul. 2002.

\bibitem{Kumar07} X. Hu, and B. V. K. V. Kumar, ``Evaluation of low-density
parity-check codes on perpendicular magnetic recording
model,"\textit{IEEE Transactions on Magnetics}, vol. 43, no. 2,
pp. 727-732, Feb. 2007.

\bibitem{Hongxin01} Hongxin Song, R. M. Todd, and J. R. Cruz, ``Applications of low-density
parity-check codes to magnetic recording channels," \textit{IEEE
Journal on Selected Areas in Communications}, vol. 19, no. 5, pp.
918-923, May 2001.

\bibitem{MoonICC05} J. Moon, and J. Park, ``Detection of prescribed error
 events: Application to perpendicular recording," in \textit{Proc. IEEE ICC}, vol. 3,
 pp. 2057-2062, May 2005.

\bibitem{intermag06} J. Park, and J. Moon, ``High-rate error correction codes targeting
dominant error patterns," \textit{IEEE Trans. Magn.}, vol. 42, no.
10, pp. 2573-2575, Oct. 2006.

\bibitem{intermag07}  J. Park, and J. Moon, ``A new class of error-pattern-correcting codes
capable of handling multiple error occurrences," \textit{IEEE
Trans. Magn.}, vol. 43., no. 6, pp. 2268-2270, Jun. 2007.

\bibitem{Jih_Moon}  J. Park, and J. Moon, ``Error-pattern-correcting cyclic codes tailored to a prescribed set of error cluster patterns," \textit{IEEE Trans.
Inform. Theory}, vol. 55, no. 4, pp. 1747 - 1765, Apr. 2009.

\bibitem{Hakim07}  H. AlHussien, J. Park and J. Moon, ``Iterative decoding based on
error-pattern correction," \textit{IEEE Trans. Magn.}, vol. 44,
no. 1, pp. 181-186, Jan. 2008.

\bibitem{wolf06}  J. K. Wolf, ``An introduction to tensor product codes and
applications to digital storage systems," \textit{Information
Theory Workshop, 2006. ITW '06 Chengdu. IEEE }, pp. 6-10, 22-26
Oct. 2006.

\bibitem{wolf65}  J. Wolf, ``On codes derivable from the tensor product of check
matrices," \textit{IEEE Transactions on Information Theory}, vol.
11, no. 2, pp. 281-284, Apr. 1965.

\bibitem{Chai06_L}  P. Chaichanavong, and P. H. Siegel, ``Tensor-product parity code
for magnetic recording," \textit{IEEE Transactions on Magnetics},
vol. 42, no. 2, pp. 350-352, Feb. 2006.

\bibitem{Chai06_P}  P. Chaichanavong, and P. H. Siegel, ``Tensor-product Parity codes:
Combination with constrained codes and application to
perpendicular recording," \textit{IEEE Transactions on Magnetics},
vol. 42, no. 2, pp. 214-219, Feb. 2006.

\bibitem{wolf63}  J. Wolf, and B. Elspas, ``Error-locating codes -- A new concept in
error control," \textit{IEEE Transactions on Information Theory},
vol. 9, no. 2, pp. 113-117, Apr. 1963.

\bibitem{Fahrner04}  A. Fahrner, H. Grie$\ss$er, R. Klarer, and V. V. Zyablov, ``Low-complexity GEL codes for digital magnetic storage
systems," \textit{IEEE Transactions on Magnetics}, vol. 40, no. 4,
pp. 3093-3095, Jul. 2004.

\bibitem{Imai81}  H. Imai, and H. Fujiya, ``Generalized tensor product codes,"
\textit{IEEE Transactions on Information Theory}, vol. 27, no. 2,
pp. 181-187, Mar. 1981.

\bibitem{Pyndiah98}  R. M. Pyndiah, ``Near-optimum decoding of product codes: Block
turbo codes," \textit{IEEE Transactions on Communications}, vol.
46, pp. 1003-1010, Aug. 1998.

\bibitem{Arnold01}   D. Arnold, and H. A. Loeliger, ``On the information rate of binary-input channels with memory,"
in \textit{Proc. IEEE Int. Conf. Communications 2001}, Helsinki,
Finland, Jun. 2001.

\bibitem{Davey98}  M. C. Davey, and D. J. C. MacKay, ``Low density parity check codes over
$GF(q)$," \textit{Information Theory Workshop}, 1998, pp. 70-71,
22-26 Jun. 1998.

\bibitem{Hu05}  X. Y. Hu, E. Eleftheriou, and D. M. Arnold, ``Regular and irregular
progressive edge-growth tanner graphs," \textit{IEEE Transactions
on Information Theory}, vol. 51, no. 1, pp. 386-398, Jan. 2005.

\bibitem{Li04}  L. Zongwang, and B. V. K. V. Kumar, ``A class of good quasi-cyclic
low-density parity check codes based on progressive edge growth
graph," \textit{Conference Record of the Thirty-Eighth Asilomar
Conference on Signals, Systems and Computers, 2004}, pp. 1990-1994
vol. 2, 7-10 Nov. 2004.

\bibitem{Chang08}  W. Chang, and J. R. Cruz, ``Performance and decoding complexity of nonbinary
 LDPC codes for magnetic recording," \textit{IEEE Trans.
Magn.}, vol. 44, no. 1, pp. 211-216, Jan. 2008.

\bibitem{Chang4k08}  W. Chang, and J. R. Cruz, ``Nonbinary LDPC codes for 4-kB sectors,"
\textit{IEEE Trans. Magn.}, vol. 44, no. 11, pp. 3781-3784, Nov.
2008.

\bibitem{ChangCom08} W. Chang, and J. R. Cruz, ``Comments on ``Performance and decoding
complexity of nonbinary LDPC codes for magnetic recording" [Jan.
08 211-216]," \textit{Magnetics, IEEE Transactions on}, vol. 44,
no. 10, pp. 2423-2424, Oct. 2008.

\bibitem{Hongxin03}  S. Hongxin, and J. R. Cruz, ``Reduced-complexity decoding of Q-ary
LDPC codes for magnetic recording," \textit{IEEE Trans. Magn.},
vol. 39, no. 2, pp. 1081-1087, Mar. 2003.

\bibitem{Lidl94} R. Lidl, and H. Niederreiter, ``Introduction to finite fields and
their applications," 2nd ed. New York: Cambridge University Press,
1994.
\end{thebibliography}
\end{document}